\documentclass[aps,prb,twocolumn,raggedbottom,floatfix]{revtex4}
\usepackage{amsmath}
\usepackage{amssymb}
\usepackage{graphicx}
\usepackage{bm}
\usepackage[dvips]{color}
\usepackage{times}
\usepackage{mathptmx}

\newcommand{\eref}[1]{eqn.\ (\ref{#1})}
\newcommand{\Eref}[1]{Eqn.\ (\ref{#1})}
\newcommand{\erefs}[2]{eqns.\ (\ref{#1},\ref{#2})}

\newcommand{\erefss}[3]{eqns.\ (\ref{#1},\ref{#2},\ref{#3})}

\newcommand{\erefseq}[2]{eqns.\ (\ref{#1} - \ref{#2})}
\newcommand{\frefseq}[2]{figs.\ (\ref{#1} - \ref{#2})}

\newcommand{\fref}[1]{fig.\ \ref{#1}}
\newcommand{\Fref}[1]{Fig.\ \ref{#1}}
\newcommand{\frefs}[2]{figs.\ (\ref{#1},\ref{#2})}
\newcommand{\frefss}[3]{figs.\ (\ref{#1},\ref{#2},\ref{#3})}
\newcommand{\sref}[1]{sec.\ \ref{#1}}
\newcommand{\srefs}[2]{secs.\ (\ref{#1},\ref{#2})}
\newcommand{\Sref}[1]{Sec.\ \ref{#1}}

\newcommand{\viz}{\emph{viz.\ }}
\newcommand{\ie}{\emph{i.e.\ }}
\newcommand{\eg}{\emph{e.g.\ }}

\newcommand{\pd}{\phantom{\dagger}}
\newcommand{\Hdot}{\hat{H}_{\mathrm{D}}}
\newcommand{\Hl}{\hat{H}_{\mathrm{L}}}
\newcommand{\Ht}{\hat{H}_{\mathrm{T}}}

\newcommand{\vsd}{V_{\mathrm{sd}}}
\newcommand{\vg}{V_{\mathrm{g}}}
\newcommand{\dvg}{\Delta V_{\mathrm{g}}}
\newcommand{\I}{\mathrm{i}}
\newcommand{\half}{\tfrac{1}{2}}
\newcommand{\K}{\mathbf{k}}

\newcommand{\ek}{\epsilon_{\mathbf{k}}^{\pd}}
\newcommand{\acre}{a^{\dagger}_{\mathbf{k} \nu \sigma}}

\newcommand{\ades}{a^{\phantom\dagger}_{\mathbf{k} \nu \sigma}}
\newcommand{\adesL}{a^{\phantom\dagger}_{\mathbf{k} L \sigma}}
\newcommand{\adesR}{a^{\phantom\dagger}_{\mathbf{k} R \sigma}}
\newcommand{\ccre}{c^{\dagger}_{\mathbf{k}\sigma}}
\newcommand{\cdes}{c^{\phantom\dagger}_{\mathbf{k} \sigma}}
\newcommand{\dicre}{d^{\dagger}_{i\sigma}}
\newcommand{\dides}{d^{\phantom\dagger}_{i\sigma}}
\newcommand{\nis}{\hat{n}_{i\sigma}^{\pd}}
\newcommand{\nic}{\hat{n}_{i}^{\pd}}
\newcommand{\vi}{\mathrm{V}_{i}^{\pd}}
\newcommand{\vin}{\mathrm{V}_{i\nu}^{\pd}}
\newcommand{\viL}{\mathrm{V}_{iL}^{\pd}}
\newcommand{\viR}{\mathrm{V}_{iR}^{\pd}}
\newcommand{\vone}{\mathrm{V}_{1}^{\pd}}
\newcommand{\vtwo}{\mathrm{V}_{2}^{\pd}}
\newcommand{\up}{U^{\prime}}
\newcommand{\J}{J_{\mathrm{H}}}

\newcommand{\Jtil}{\tilde{J}_{\mathrm{H}}}
\newcommand{\util}{\tilde{U}}
\newcommand{\uptil}{\tilde{\up}}
\newcommand{\etilone}{\tilde{\epsilon}_{1}}
\newcommand{\etiltwo}{\tilde{\epsilon}_{2}}

\newcommand{\nimp}{n_{\mathrm{imp}}}
\newcommand{\simp}{S_{\mathrm{imp}}(T)}
\newcommand{\Simp}{S_{\mathrm{imp}}}
\newcommand{\ximp}{\chi_{\mathrm{imp}}(T)}
\newcommand{\Ximp}{\chi_{\mathrm{imp}}}

\begin{document}

\title{Correlated electron 
physics in 
multilevel quantum dots: \\ phase transitions, transport, and experiment.}
\date{\today}
\author{David E. Logan}
\author{Christopher J. Wright}
\author{Martin R. Galpin}
\affiliation{Oxford University, Chemistry Department, Physical \& Theoretical Chemistry, 
South Parks Road, Oxford, OX1~3QZ, UK.}

\begin{abstract}
We study correlated two-level quantum dots, coupled in effective 1-channel fashion to metallic leads; with electron interactions including on-level and inter-level Coulomb repulsions, as well as the inter-orbital Hund's rule exchange favoring the spin-1 state in the relevant sector of the free dot. For arbitrary dot occupancy,
the underlying phases, quantum phase transitions (QPTs), thermodynamics, single-particle dynamics and electronic transport properties are considered; and direct comparison is made to conductance experiments on lateral quantum dots. Two distinct phases arise generically, one characterised by a normal Fermi liquid fixed point (FP), the
other by an underscreened (USC) spin-1 FP. Associated QPTs, which occur in general in a mixed valent regime of non-integral dot charge, are found to consist of continuous lines of Kosterlitz-Thouless transitions, separated by first order level-crossing transitions at high symmetry points. A `Friedel-Luttinger sum rule' is derived and,
 together with a deduced generalization of Luttinger's theorem to the USC phase (a singular Fermi liquid),
is used to obtain a general result for the $T=0$ zero-bias conductance, expressed solely in terms of the dot occupancy and applicable to both phases. Relatedly, dynamical signatures of the QPT show two broad classes of behavior, corresponding to the collapse of either a Kondo resonance, or antiresonance, as the transition is approached from the Fermi liquid phase; the latter behavior being apparent in experimental differential conductance maps. The problem is studied using the numerical renormalization group method, combined with analytical arguments.
\end{abstract}


\maketitle

\section{Introduction}
\label{sec:intro}
The Kondo effect is one of the enduring paradigms of quantum many-body theory~\cite{Hewsonbook}. 
For most of its history it has been associated with bulk condensed matter, notably transition metal impurites dissolved in clean metals, and certain heavy fermion rare earth compounds~\cite{Hewsonbook}.
In recent years, however, the advent of quantum dot systems -- with the impressive control and tunability 
possible for `artificial atoms' -- has generated a strong resurgence of interest in Kondo and related physics in nanoscale devices (for reviews see \eg [\onlinecite{kouwenhovenqdreview,PustGlazRev}]). 

  In odd-electron quantum dots the spin-$1/2$ Kondo effect arises. Manifest experimentally~\cite{goldhaber,cronenwett} as a strong low-temperature enhancement of the zero bias conductance, 
indicating the formation of the local Kondo singlet below a characteristic Kondo temperature,
the basic theoretical model here is of course the Anderson impurity model~\cite{anderson}: a single dot level, with a single on-level Coulomb interaction, tunnel coupled to non-interacting metallic leads. Moreover 
the Anderson model captures not only the Kondo regime --  arising towards the center of the  associated Coulomb blockade valley where the dot level is singly occupied  -- but also the mixed valent regimes of non-integral occupancy occurring towards the edges of the valley. As such, it encompasses essentially all the physics associated with a single `active' dot level.

  The situation is naturally more complex, and richer, if two active dot levels are integral to electronic transport. For example, higher dot spin states now become possible, in this case a 2-electron triplet stabilised by the inter-orbital Hund's rule exchange~\cite{Brouwer1999,Baranger2000}. This state has been observed experimentally in even-electron dots, for both  lateral~\cite{Schmid2000,VanderWielPRL2002,Kogan,Quay2007}
and vertical~\cite{SasakiNature2000} devices (as well as in a single-molecule dot~\cite{RochFlorens2008}).
It too is manifest in a strong enhancement of the zero-bias conductance, indicative~\cite{PustGlaz2001,Posazh2006} of proximity to an underscreened spin-1 fixed point~\cite{nozblan} in which the spin-$1$ is quenched to an effective spin-$\half$ on coupling to the leads. 

Much important theoretical work on the problem has ensued; including both the 1-channel case 
(see \eg [\onlinecite{Kikoin2001,HofSch,vojtabullahof02,KollerHewson,pustborda,Posazh2005,Posazh2006,AligiaPreprint2009}])
where the single screening channel yields underscreened (USC) spin-$1$ as the stable low-temperature fixed point,
and the 2-channel case~\cite{PustGlaz2000,PustGlaz2001,PustGlazHof2003,HofZarand2004,Posazh2005,Posazh2006} where the spin-$1$ local moment is fully screened at the lowest temperatures. 
Further, since the USC spin-1 fixed point is clearly distinct from that characteristic of a normal Fermi liquid
-- the USC phase being a `singular Fermi liquid'~\cite{mehta} -- quantum phase transitions from a normal Fermi liquid to the USC phase are expected, and found, to arise in the 1-channel case (with pristine transitions broadened into crossovers for 2-channel screening).  This too has been studied quite extensively~\cite{Kikoin2001,HofSch,vojtabullahof02,pustborda,AligiaPreprint2009,PustGlaz2000,PustGlazHof2003,HofZarand2004}. However the large majority of previous work on these `singlet-triplet' transitions has focussed on a 
somewhat particular case -- the middle of the 2-electron Coulomb blockade valley where, throughout both phases, the dot occupancy/charge remains close to 2; a situation we regard as unlikely to be applicable to a transition driven by tuning a gate voltage in the absence of a magnetic field (as in the experiments \eg of [\onlinecite{Kogan}]), where one instead expects the dot charge to vary continuously with gate voltage. A notable exception 
is the work of [\onlinecite{pustborda}], in which low-temperature transport is considered in a region separating two adjacent Coulomb blockade valleys with spins $S=\half$ and $S=1$ on the dot, and where the resultant quantum phase transition, driven by gate voltage and arising in the limit $B\rightarrow 0+$ of vanishing magnetic field, occurs in a mixed valent regime of non-integral dot charge.

  In view of the above our aim here is to consider a rather general model of a two-level quantum dot, coupled in
a 1-channel fashion to metallic leads; to consider its underlying phases, thermodynamics, single-particle dynamics and associated low-temperature ($T$) electronic transport, for \emph{arbitrary} dot charge -- spanning as such the full range of possible behavior;  and ultimately to make tangible comparison to experiment~\cite{Kogan}. The model itself is specified in  \sref{sec:model} and reflects the natural complexity
of a two-level dot, where in addition to the one-electron dot levels electron interactions include both on-level and inter-level Coulomb repulsions, together with inter-orbital spin-exchange. We study it using Wilson's numerical renormalization group (NRG) technique~\cite{wilson,kww1,kww2} as the method of choice, employing the full density matrix formulation of the method~\cite{PetersPruschkeAnders2006,WeichselbaumDelft2007} (for a recent review see
[\onlinecite{NRGrmp}]); together where possible with analytical arguments. 

The intrinsic phases and associated thermodynamics are considered in \sref{sec:thermo}. With $\epsilon_{1}$,
$\epsilon_{2}$ denoting the one-electron level energies, the general structure of the phase diagrams in the ($\epsilon_{1},\epsilon_{2}$)-plane is found to consist of a closed, continuous line of quantum phase transitions (QPTs) separating an USC spin-1 phase from a continuously connected normal Fermi liquid phase; 
 although more complex topologies arise as the exchange coupling is driven weakly antiferromagnetic (\sref{sec:AFJ}), leading ultimately to destruction of the USC phase. The transitions are found in general to be of Kosterlitz-Thouless type, except for particular lines of symmetry where first order, level-crossing transitions arise (\sref{sec:firstorderthermo}).

\Sref{sec:dynamics} focusses on the $T=0$ zero-bias conductance $G_{c}$, and associated static phase shift
$\delta$. A `Friedel-Luttinger sum rule' for $\delta$ is derived, applicable to both the normal Fermi liquid  
and the USC spin-1 phases, and reducing to the usual Friedel sum rule~\cite{Langreth1966,Hewsonbook} in the Fermi liquid phase. Since the USC phase is a singular Fermi liquid~\cite{mehta}, and as such not perturbatively connected to the non-interacting limit of the model, one does not expect Luttinger's (integral) theorem~\cite{luttingerfs} to
apply. A generalization of  it for the USC phase is however deduced, and its important consequences for the zero-bias conductance considered; leading to a simple result which, for \emph{both} the normal Fermi liquid and USC  phases, gives $G_{c}$ in terms of the dot occupancy/charge (or, strictly, the `excess impurity charge'~\cite{Hewsonbook}).

  Single-particle dynamics for both phases are detailed in \sref{sec:spectra}. In particular, dynamical signatures of the QPT on approaching it from the normal Fermi liquid are found to fall into two broad classes, corresponding respectively to an `on the spot' vanishing of either a Kondo resonance, or a Kondo antiresonance, in the single-particle spectrum; the spectral collapse in either case being associated with a vanishing Kondo scale $T_{K}$ as the transition is approached, and in terms of which universal scaling of dynamics is found to occur.

  Finally, in \sref{sec:exp} we make explicit comparison to the experiments of [\onlinecite{Kogan}] on a lateral dot; in which, on continuous tuning of a gate voltage at zero magnetic field, both the normal spin-$1/2$ Fermi liquid and the USC spin-1 phase are observed in adjacent Coulomb blockade valleys. Both the zero-bias conductance as a function of  gate voltage, and (in this case inevitably approximate) differential conductance maps as a function of both gate and bias voltages, are compared to experiment; and the features observed related to the dynamics considered in \srefs{sec:dynamics}{sec:spectra}. We believe it fair to say that the underlying theory accounts rather well for experiment.
  

\section{Model}
\label{sec:model}

Interacting quantum dots and other nanodevices are described generally by the dot $\Hdot$, a pair of non-interacting leads $\Hl$, and a tunnel coupling between the subsystems: $\hat{H}=\Hdot +\Hl +\Ht$. We consider 
in this work a two-level interacting quantum dot of form:
\begin{equation}
\label{eq:hdot}
\Hdot~=~\sum_{i,\sigma} \left( \epsilon_{i}+\half U\hat{n}_{i-\sigma}^{\pd}\right)\nis +\up\hat{n}_{1}^{\pd}\hat{n}_{2}^{\pd}
-\J~\hat{\mathbf{s}}_{1}\cdot\hat{\mathbf{s}}_{2}
\end{equation}
Here $\nis =\dicre\dides$ where $\dicre$ creates a $\sigma$ ($=\uparrow,\downarrow$) spin electron in level $i$ ($=1,2$), $\nic =\sum_{\sigma} \nis$ is the total number operator for level $i$, and $\hat{\mathbf{s}}_{i}$ is the local spin-operator with components 
$\hat{\mathbf{s}}_{i}^{\alpha} =\sum_{\sigma,\sigma^{\prime}}\dicre
\sigma^{(\alpha)}_{\sigma\sigma^{\prime}}d_{i\sigma^{\prime}}^{\pd}$ and $\bm{\sigma}^{(\alpha)}$
the Pauli matrices. The single-particle levels have energies $\epsilon_{i}$, the on-level Coulomb interaction 
(taken to be the same for both levels) is denoted by $U$, and the inter-level interaction by $\up$.
Finally $\J$ is the exchange coupling, taken in accordance with Hund's rule to be ferromagnetic
 ($\J >0$, although we also comment in \sref{sec:AFJ} on the weakly antiferromagnetic case). The 
states arising from $\Hdot$ itself will be discussed in \sref{sec:phaseoverview} below.

\begin{figure}
\includegraphics[scale=0.8]{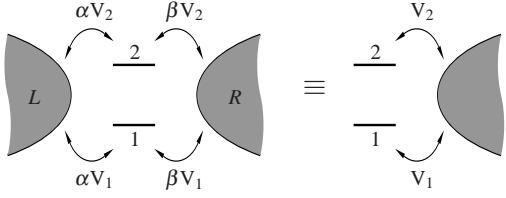}
\caption{\label{fig:fig1} Schematic of the two-level dot (levels 1,2) coupled to the $\nu =L,R$ leads with
the tunnel couplings  $V_{2\nu}/V_{1\nu} \equiv V_{2}/V_{1}$ independent of $\nu$; leading to
the equivalent one-lead description indicated.
}
\end{figure}

  The Hamiltonian for the two equivalent non-interacting leads ($\nu = L,R$) is given by
$\Hl=\sum_{\nu}\sum_{\K,\sigma}\ek \acre\ades$. Tunnel coupling to the leads is described generally
by $\Ht=\sum_{\nu}\sum_{i,\K,\sigma}\vin (\dicre\ades + \mathrm{h.c.})$ where $\vin$ is the tunnel coupling matrix element between dot level $i$ and lead $\nu$. We consider explicitly in this paper the case of an effective 1-channel setup, in which the ratio $V_{2\nu}^{\pd}/V_{1\nu}^{\pd} \equiv \vtwo/\vone$ is independent of the lead
index $\nu$; \ie the tunnel couplings are of form  $\viL =\alpha \vi$, $\viR=\beta\vi$ (with $\alpha^{2}+\beta^{2} =1$), as illustrated schematically in \fref{fig:fig1}.
A simple canonical transformation to new lead orbitals may then be performed,
$\cdes =\alpha\adesL +\beta\adesR$ and $\tilde{c}_{\K\sigma}^{\phantom\dagger}=-\beta\adesL +\alpha\adesR$,
such that solely the bonding combination of lead states ($\cdes$) couples to the dot:
\begin{equation}
\label{eq:ht}
\Ht~=~\sum_{i,\K,\sigma}\vi (\dicre\cdes +\mathrm{h.c.})
\end{equation}
We can thus drop the lead index $\nu$ and consider one effective lead
\begin{equation}
\label{eq:hl}
\Hl ~=~\sum_{\K,\sigma}\ek\ccre\cdes~,
\end{equation}
hence the effective 1-channel description illustrated in \fref{fig:fig1}. In practice we consider the standard case~\cite{Hewsonbook} of a symmetric, flat-band conduction band  with half bandwidth $D$, \ie the lead density of states (per conduction orbital) is $1/(2D)$.

It need hardly be added that the tunnel coupling pattern considered (\fref{fig:fig1}) is not the 
most general case, which would by contrast involve an irreducibly 2-channel description~\cite{PustGlaz2001}, in general with strong channel anisotropy~\cite{PustGlaz2001,Posazh2006}. The richness of the physics arising in the case considered, with its associated pristine QPTs, is nonetheless more than ample to justify its study (indeed for NRG calculations in practice, we focus largely on the case $\vtwo =\vone$).
It has moreover been argued (see \eg [\onlinecite{HofSch,vojtabullahof02}]) that the 1-channel case 
is generally appropriate to lateral quantum dots, while a 2-channel model is appropriate for vertical dots.

In considering equilibrium electronic transport \emph{per se}, the central quantity is 
of course the zero-bias conductance, $G_{c}(T)$, across the $L/R$ leads (\fref{fig:fig1}).
An expression for it is readily obtained following Meir and Wingreen~\cite{meirwingreen}, 
and is:
\begin{equation}
\label{eq:zbc}
G_{\mathrm{c}}(T)~=~\frac{2e^{2}}{h} G_{0}\int^{\infty}_{-\infty} d\omega ~\frac{-\partial f(\omega)}{\partial \omega}~ \pi (\Gamma_{11} +\Gamma_{22})~D_{ee}(\omega)
\end{equation}
Here $f(\omega) = [e^{\omega/T}+1]^{-1}$ ($k_{B} \equiv 1$) is the Fermi function and
\begin{equation}
\label{eq:gammaii}
\Gamma_{ii}~=~\pi \rho \mathrm{V}_{i}^{\pd 2}
\end{equation}
is the hybridization strength for level $i$ ($\rho$ is the lead density of states).
The dimensionless conductance prefactor $G_{0}=(2\alpha\beta)^{2}$ -- or equivalently
$G_{0}=4\Gamma_{L}\Gamma_{R}/(\Gamma_{L}+\Gamma_{R})^{2}$ with
$\Gamma_{\nu}=\pi\rho \sum_{i}\mathrm{V}_{i\nu}^{\pd 2}$ -- reflects the relative asymmetry in tunnel coupling
to the $L/R$ leads. It is naturally maximal, $G_{0}=1$, for symmetric coupling where (\fref{fig:fig1}) $\alpha = 1/\sqrt{2} =\beta$ (\ie $\Gamma_{R}=\Gamma_{L}$). The key quantity determining the conductance \eref{eq:zbc}, which we analyse in detail in later sections, is  the `even-even' single-particle spectrum: 
$D_{ee}(\omega)=-\tfrac{1}{\pi}\mathrm{Im}G_{ee}(\omega)$ in terms of the (retarded) Green function
$G_{ee}(\omega)$ ($\leftrightarrow G_{ee}(t)=-\I \theta(t)\langle \{d_{e\sigma}^{\pd}(t),d_{e\sigma}^{\dagger}\}\rangle$). The $e$-orbital creation operator is given 
generally by
\begin{equation}
\label{eq:evenorb}
d_{e\sigma}^{\dagger}~=~ \frac{1}{\sqrt{\Gamma_{11}+\Gamma_{22}}} \left( \sqrt{\Gamma_{11}}~
d_{1\sigma}^{\dagger} ~+~\sqrt{\Gamma_{22}}~d_{2\sigma}^{\dagger}\right)
\end{equation}
in terms of the level creation operators, such that
\begin{equation}
\label{eq:gee}
G_{ee}(\omega)~=~ \frac{1}{\Gamma_{11}+\Gamma_{22}}~\sum_{i,j}~ \Gamma_{ij}~G_{ij}(\omega)
\end{equation}
in terms of the corresponding propagators for the dot levels, $G_{ij}(\omega)$ ($i,j \in{\{1,2\}}$); and
where
\begin{equation}
\label{eq:gamma12}
\Gamma_{12}~=~\pi\rho\vone\vtwo ~~~ \equiv \sqrt{\Gamma_{11}\Gamma_{22}}
\end{equation}
is the inter-level hybridization strength.

For the case $\vtwo =\vone$ all hybridization strengths coincide,
\begin{equation}
\label{eq:equalgammas}
 \Gamma_{ij}~ \equiv ~\Gamma ~~~~~~~~: \vtwo =\vone
\end{equation}
(and the $e$-$e$ propagator then reduces simply to 
$G_{ee}(\omega) =\half [G_{11}(\omega)+G_{22}(\omega)+2G_{12}(\omega)]$). It is convenient
in this case to specify the `bare' parameters of $\Hdot$ in terms of $\Gamma$, defining
\begin{equation}
\label{eq:tildes}
\tilde{\epsilon_{i}} = \frac{\epsilon_{i}}{\Gamma}, ~~~~~\tilde{U} = \frac{U}{\Gamma},
~~~~~\tilde{\up} = \frac{\up}{\Gamma}, ~~~~~\Jtil = \frac{\J}{\Gamma}~.
\end{equation}


\subsection{Symmetries}
\label{sec:symmetries}

We will subsequently consider different phases of the dot-lead coupled system in the 
($\epsilon_{1},\epsilon_{2}$)-plane, for given values of the interaction parameters $U,\up$ and $\J$
entering $\Hdot$  (\eref{eq:hdot}). To this end it is economical to exploit symmetry.
Rather than the bare levels $\epsilon_{1},\epsilon_{2}$ it is often helpful 
to employ 
\begin{subequations}
\label{eq:xandy}
\begin{equation}
\label{eq:x}
x ~=~ \epsilon_{1}~+~\half U~+~\up
\end{equation}
\begin{equation}
\label{eq:y}
y ~=~ \epsilon_{2}~+~\half U~+~\up ~.
\end{equation}
\end{subequations}
Their significance arises from a particle-hole transformation ($p$-$ht$) of
$\hat{H}=\Hdot +\Hl +\Ht$ (\erefseq{eq:hdot}{eq:hl}), namely~\cite{kww1}
\begin{equation}
\label{eq:pht}
\dides ~ \rightarrow ~ \dicre  ~~~~~~~\cdes \rightarrow -c^{\dagger}_{-\K\sigma}~.
\end{equation}
$\hat{H} \equiv \hat{H}(x,y)$ transforms under the $p$-$ht$  as 
$\hat{H}(x,y) \rightarrow 2(x+y)+\hat{H}(-x,-y)$, and is hence invariant at the $p$-$h$ symmetric point
$x=0=y$. Use of $x, y$ thus specifies the level energies relative to this point. All physical properties,
thermodynamic and dynamic, have characteristic symmetries under the $p$-$ht$, which we exploit many 
times in the paper. For example the free energy $F(x,y) = -T \ln \{\mathrm{Tr}~e^{-\beta \hat{H}(x,y)}\}$
is,  modulo an irrelevant constant, equivalent to its $p$-$ht$ counterpart ($F(x,y)=2(x+y) +F(-x,-y)$),
whence \eg phase boundaries (\srefs{sec:phaseoverview}{sec:thermo}) are invariant under inversion 
$(x,y) \rightarrow (-x,-y)$; and thus only $y \geq x$ need in practice be considered.

  The second symmetry exploited is a `$1$-$2$' transformation, \viz the trivial canonical transformation
\begin{equation}
\label{eq:12t}
(d_{1\sigma}^{\pd},d_{2\sigma}^{\pd})~\rightarrow ~ (d_{2\sigma}^{\pd},d_{1\sigma}^{\pd})~
\end{equation}
under which the dot Hamiltonian $\Hdot(x,y) \rightarrow \Hdot(y,x)$. The same symmetry applies to the
full $\hat{H}$ for $\vtwo =\vone$; whence \eg $F(x,y)=F(y,x)$ is invariant to reflection about the
line $y=x$, and in consequence phase boundaries need overall be considered only  for $y \geq |x|$.

\begin{figure*}
\includegraphics{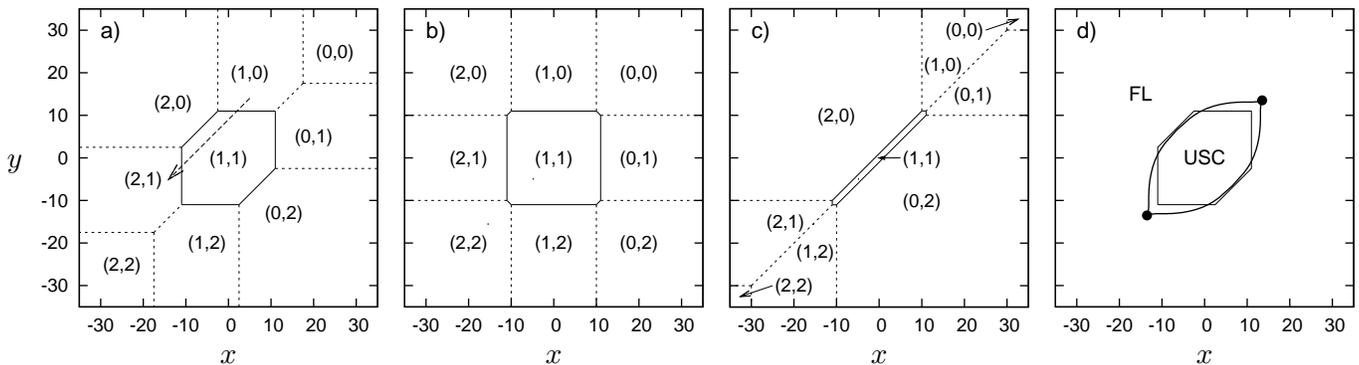}
\caption{\label{fig:fig2} (a)-(c) show ground states of the isolated dot in the ($x,y$)-plane, for $U=20$ and
$J_{\mathrm{H}}= 5$  (in units of $\Gamma \equiv 1$); with (a) $\up = 7.5$, (b) $\up = 0$ and
(c) $\up =U$. States are labelled as $(n_{1},n_{2})$, with $n_{i}=\langle\hat{n}_{i}\rangle$ the charge on level $i$.
Phase boundaries surrounding the 2-electron triplet state $(n_{1},n_{2})=(1,1)$) are indicated by solid lines, 
and all others by dotted lines. The dashed arrowed line in (a) shows the form of an experimental `trajectory' on application of a gate voltage, see text. For the same bare parameters as (a), \Fref{fig:fig2}(d) shows the phase diagram obtained via NRG for the lead-coupled dot system (with $V_{2}=V_{1}$), as detailed in \sref{sec:thermo} \emph{ff}. It consists of a line of continuous quantum phase transitions (thick solid line) and two first order level-crossing transitions on the line $y=x$ (shown as dots), separating a singular Fermi liquid phase~\cite{mehta} characterised by an underscreened spin-1 fixed point~\cite{nozblan} (interior, `USC') from a normal Fermi liquid (`FL') characterised in general by a frozen impurity FP~\cite{kww2}. The hexagonal boundary of the $(1,1)$ triplet state for the isolated dot (\fref{fig:fig2}(a)) is also shown for comparison.
}
\end{figure*}


\subsection{Ground state phases: overview}
\label{sec:phaseoverview}
It is first instructive to consider briefly the states of the
isolated dot in the ($x,y$)-plane, as determined by the ground states of $\Hdot$ (\eref{eq:hdot}). We label the
dot states as ($n_{1},n_{2}$) (with $n_{i}=\langle\nic\rangle$ the ground state charge for level $i$),
with energies $E_{D}(n_{1},n_{2})$. For all $\J >0$, the 2-electron dot ground state is the ($1,1$) spin triplet
with energy $E_{T}$ ($\equiv E_{D}(1,1) =\epsilon_{1}+\epsilon_{2}+\up -\tfrac{1}{4}\J$), centred on the $p$-$h$ symmetric point $(x,y) =(0,0)$; as illustrated in \fref{fig:fig2}(a) for the representative case $U=20$, 
$\up=7.5$ and $\J=5$. All other ground states, indicated in the figure, are either spin singlets or doublets.

Considering in particular $y\geq |x|$  -- phase boundaries being invariant to inversion and reflection as 
above -- the ($1,1$) triplet is bordered both by another 2-electron state, \viz the spin singlet ($2,0$), and by a 1-electron spin doublet ($1,0$) (the bounding lines for which are
given by $y=x+U-\up +\tfrac{1}{4}\J$ and $y=\tfrac{1}{2}U +\tfrac{1}{4}\J$ respectively).
The dashed line in \fref{fig:fig2}(a) shows a typical `trajectory', $y=x +\Delta\epsilon$
(\ie $\epsilon_{2}=\epsilon_{1}+\Delta\epsilon$), expected from experiment on application of a gate voltage $\vg$ to the dot, with $\epsilon_{1} \propto \vg$ and fixed level spacing  $\Delta\epsilon =\epsilon_{2}-\epsilon_{1}$. 
Note that the 2-electron $(1,1)$ triplet is thereby accessed from the 1-electron state $(1,0)$~\cite{pustborda} (as relevant to comparison with experiment, \sref{sec:exp}).

  \Fref{fig:fig2}(b,c) show the isolated dot ground states arising for inter-level Coulomb repulsion $\up =0$ and $\up =U$, respectively. For $y>|x|$ in the former case, the $(1,1)$ triplet state is bordered almost exclusively
by the 1-electron state $(1,0)$, while in the latter case it is bordered almost exclusively by the 2-electron singlet $(2,0)$. These two cases are of course extremes; and although aspects of the model  have
been considered previously for the case $\up =U$~\cite{Kikoin2001,HofSch,pustborda}, 
we know of no compelling reason why the intra- and inter-level Coulomb repulsions should in general be near coincident for reasonably small dots (indeed we argue in \sref{sec:exp} that comparison to the experiment of [\onlinecite{Kogan}] is consistent with the contrary).

  The states of the dot \emph{per se} are of course quite trivial. We now consider the full lead-coupled system, our aim here being to give simple qualitative arguments for the general form of the phase diagram in the ($x,y$)-plane.
  
    On coupling to the leads, the effective low-energy model deep in the spin-1 regime centred on $(x,y)=(0,0)$,
is naturally a 1-channel spin-1 Kondo model~\cite{PustGlaz2001,Posazh2006} (obtained formally by a Schrieffer-Wolff transformation~\cite{schriefferwolff,Hewsonbook} retaining only the triplet ($1,1$)-state of the dot itself,
see also Appendix A). Its low-energy physics is well known~\cite{nozblan}: half the spin-1 is screened by the conduction electrons, leading to a free spin-$\tfrac{1}{2}$ with weak residual ferromagnetic coupling to the metallic lead, which results in in non-analytic (logarithmic) corrections to Fermi liquid behavior; the resultant state being classified as a singular Fermi liquid~\cite{mehta}.  The associated low-energy fixed point (FP) is of course the underscreened spin-1 (USC) FP of Nozi\`eres and Blandin~\cite{nozblan}.

  By contrast, deep in the 1-electron ($1,0$)-regime (\fref{fig:fig2}), the effective low-energy 
model is obviously spin-$\tfrac{1}{2}$ Kondo, a normal Fermi liquid with a fully quenched spin and a strong coupling (SC) low-energy FP~\cite{Hewsonbook,wilson,kww1}. Since the underlying stable FPs (USC and SC) associated with 
these two regimes are fundamentally distinct, a quantum phase transition (QPT) somewhere between the two must 
therefore occur~\cite{pustborda}.

But what of the other isolated dot states, encircling the spin-1 state as illustrated in \fref{fig:fig2}(a-c) (and all of which as noted above are either spin singlets or doublets)? The salient point here is that, on coupling to the leads, \emph{all} such give rise to Fermi liquid states: their stable low-energy FPs form a continuous line 
connecting the SC FP arising for the spin-$\half$ Kondo model to the generic case of the frozen impurity 
FP~\cite{kww2} (as follows from the original work of Krishnamurthy, Wilkins and Wilson~\cite{kww2} on the asymmetric single-level Anderson model). No phase transitions between these states can therefore occur, the `transitions' arising in the isolated dot limit (dotted lines in \fref{fig:fig2}(a-c)) being replaced by continuous crossovers.

In consequence, one expects the general structure of the phase diagram in the ($x,y$)-plane to consist of a closed, continuous line of QPTs separating an USC spin-1 phase from a continuously connected normal Fermi liquid phase. 
This is indeed as found from detailed NRG analysis, as will be seen in the following sections. A
typical resultant phase diagram is shown in \fref{fig:fig2}(d) (for the same bare parameters as \fref{fig:fig2}(a),
the phase boundary occurring close to the border of the $(1,1)$ state of the isolated dot as one might expect).
It consists of a line of continuous QPTs; together with two first order level-crossing QPTs on the line $y=x$ (indicated by dots in \fref{fig:fig2}(d)), which are equivalent to each other under the $p$-$h$ transformation $x \rightarrow -x$. 

The transitions will be discussed in detail below, but we add here that the occurrence of first order transitions along the $y=x$ line ($\epsilon_{2}=\epsilon_{1}$) is a general consequence of symmetry. As noted in \sref{sec:symmetries}, for $V_{2}=V_{1}$ the full $\hat{H}$ transforms under the `1-2' transformation as $\hat{H}(x,y) \rightarrow \hat{H}(y,x)$, and is hence invariant on the line $y=x$. Along that line all states of the entire system thus have definite parity under the `1-2' transformation, with the Hilbert space of $\hat{H}$ strictly
separable into disjoint parity sectors. A level-crossing transition must thus occur when the global many-body ground state changes parity (further discussion of it will be given below).


\section{Phases and thermodynamics}
\label{sec:thermo}

 Dynamics and transport properties will be discussed in \sref{sec:dynamics} \emph{ff}, 
but we begin with thermodynamics; in particular the temperature ($T$) dependence of 
two standard quantities~\cite{Hewsonbook,NRGrmp} which provide clear signatures of the various FPs reached under renormalization on decreasing the temperature/energy scale, namely the entropy $\simp$ and the uniform spin susceptibility $\ximp =\langle (\hat{S}^{z})^{2}\rangle_{\mathrm{imp}}/T$ (where $\hat{S}^{z}$ refers to the spin of the entire system, and $\langle\hat{\Omega}\rangle_{\mathrm{imp}} = \langle\hat{\Omega}\rangle -\langle\hat{\Omega}\rangle_{0}$ with $\langle\hat{\Omega}\rangle_{0}$ denoting a thermal average in the absence of the dot).

We also consider briefly the usual $T=0$ `excess impurity charge' $\nimp$, \viz the difference in charge of the entire system with and without the dot present (${\hat\Omega}\equiv \hat{N}=\sum_{\K ,\sigma}\ccre\cdes + \sum_{i}\nic$ in the above); and which in practice corresponds closely to the net dot charge,
$\nimp \simeq  \langle \hat{n}_{1}+\hat{n}_{2}\rangle$, see also \sref{sec:GFs}. Prosaic though $\nimp$ is, we show later that it plays a key role in understanding the 
zero-bias conductance in \emph{both} the USC and FL phases, and relatedly the `Friedel-Luttinger sum rule' of \sref{sec:FLsumrule}. Under the $p$-$h$ and 1-2 transformations of \sref{sec:symmetries}, $\nimp \equiv \nimp(x,y)$ transforms respectively as:
\begin{subequations}
\label{eq:nimpsymm}
\begin{align}
\nimp(x,y)~&=~4-\nimp(-x,-y)
\label{eq:nimpp-hsym}\\
&=~\nimp(y,x)
\label{eq:nimp1-2sym}
\end{align}
\end{subequations}

Results shown are obtained using the full density matrix
formulation~\cite{PetersPruschkeAnders2006,WeichselbaumDelft2007} of Wilson's non-perturbative NRG 
technique~\cite{wilson,kww1,kww2}, employing a complete basis set of the Wilson chain; for a recent review see~[\onlinecite{NRGrmp}]. Calculations are typically performed for an NRG discretization parameter $\Lambda =3$, retaining the lowest 2000 states per iteration. We here consider explicitly the case $V_{2}=V_{1}$ (\sref{sec:model}), with the hybridization $\Gamma$ (\eref{eq:equalgammas}) as the basic energy unit, choosing the lead bandwidth $D/\Gamma =100$ ($\gg1$, such that results are  independent of $D$ for all practical purposes).

\begin{figure}
\includegraphics{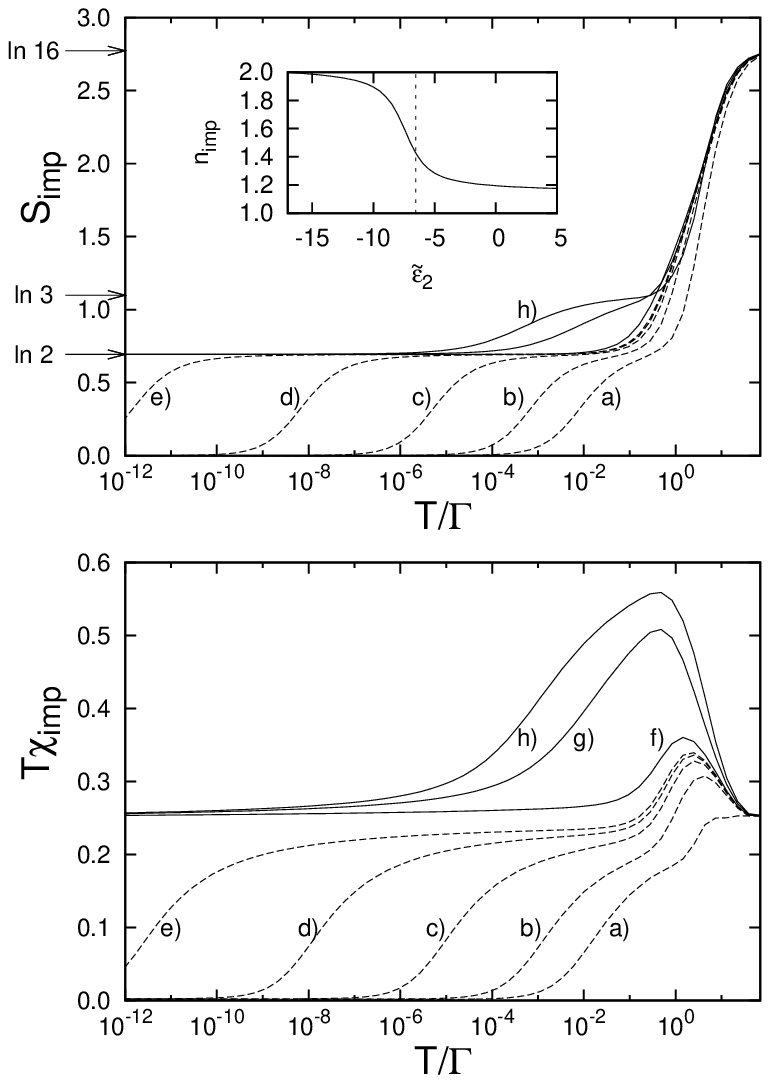}
\caption{\label{fig:fig3} $T$-dependence of the entropy $\simp$ (top panel) and spin susceptibility
$T\ximp$ (bottom); for fixed level energy $\etilone =-\tfrac{1}{2}\util -\uptil$ (\ie $x=0$) on progressively
decreasing $\etiltwo$ from deep in the FL phase, across the QPT (at 
$\tilde{\epsilon}_{2c} = -6.536..$), and through to the $p$-$h$ symmetric point atthe center of 
the USC phase. Shown for $\util =20$, $\uptil =7$ (\ie $\etilone = -17$) and $\tilde{J}_{\mathrm{H}} =2$,
with: $\etiltwo =$ +$5$ (a), $-5$ (b), $-6$ (c), $-6.3$ (d), $-6.43$ (e), $-7$ (f), $-10$ (g) and $-17$ (h). 
[Labels f) and g) are omitted for clarity from the top panel, but are easily identified from the bottom.] ~
\emph{Inset, top panel}: the corresponding $T=0$ impurity charge $\nimp$ \emph{vs} $\etiltwo$. It changes continuously as the QPT is crossed at $\tilde{\epsilon}_{2c}$ (dashed vertical line), at which point $\nimp \simeq 1.4$.
}
\end{figure}

  \Fref{fig:fig3} shows the $T/\Gamma$-dependence of $\simp$ (top) and $T\ximp$ (bottom), for fixed
$\util =20$, $\uptil = 7$ and $\Jtil =2$, taking a vertical cut through the $(x,y)$-phase diagram: the energy of
level-1 is fixed at $\etilone = -\half\util -\uptil \equiv -17$ (\ie $x=0$), and $\etiltwo$ (or equivalently 
$y$) is progressively decreased from deep in the FL phase, towards and through the transition, and down to the
$p$-$h$ symmetric point at the center of the USC phase; the transition occurring at
$\tilde{\epsilon}_{2c} = -6.536...$ (close to the value $-6.5$ expected from the isolated dot limit).

\begin{figure}
\includegraphics{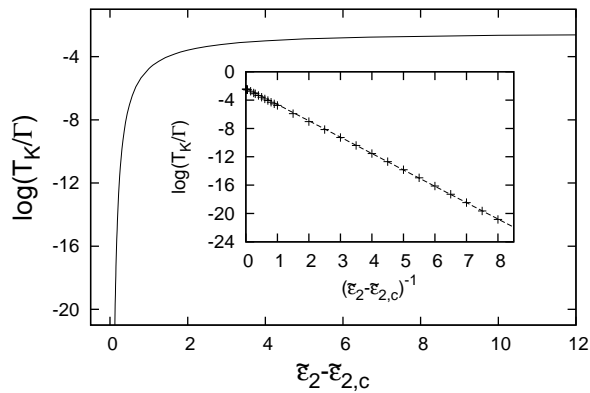}
\caption{\label{fig:fig4} Evolution of the Kondo scale in the FL phase
$\tilde{\epsilon}_{2} >\tilde{\epsilon}_{2c}$ for the same parameters as \fref{fig:fig3}: 
$\log (T_{K}/\Gamma)$ \emph{vs} $\tilde{\epsilon}_{2}-\tilde{\epsilon}_{2c}$.
Inset: $T_{K}$ vanishes exponentially as the QPT is approached, $T_{K} \propto \exp{(-a/|\tilde{\epsilon}_{2}-\tilde{\epsilon}_{2c}|)}$ 
(with $a \sim {\cal{O}}(1)$), 
characteristic of a Kosterlitz-Thouless transition.
}
\end{figure}

In all cases the highest $T$ behavior is naturally governed by the free orbital FP~\cite{kww1,kww2}, 
with all $4^{2}$ states of the two-level dot thermally accessible, hence $\Simp =\ln 16$
(and $T\Ximp = 2 \times \tfrac{1}{8}$). For case (a), $\etiltwo =+5$ is sufficiently large that level-2 is in essence irrelevant (provided $T/\Gamma \ll \etiltwo -\etilone$), the model thus reducing in effect to a
\emph{single}-level Anderson model~\cite{kww1,kww2}. Hence, on decreasing $T$, $\simp$ first flows towards
the spin-1/2 local moment (LM) FP corresponding to $\Simp =\ln 2$ (evident in this case
as a relatively weak plateau at $T/\Gamma \lesssim 1$, reflecting the modest minimum thermal
excitation of $\sim E_{D}(2,0)-E_{D}(1,0) = \epsilon_{1}+U =3\Gamma$). On further decreasing $T$, the system
then flows to the stable frozen impurity (FI) FP symptomatic of the Fermi liquid ground state, with vanishing
entropy $\Simp$ (likewise $T\Ximp =0$). A Kondo scale $T_{K}$ may be identified from the crossover between the marginally unstable LM FP  and the stable FI FP (we define it in practice via $S_{\mathrm{imp}}(T_{K}) = 0.1$).
On further decreasing $\etiltwo$, cases (b-e) in \fref{fig:fig3}, the same essential behavior is found,
the FI FP remaining the stable low-$T$ FP. But the $\Simp = \ln 2$ (and $T\Ximp \simeq \tfrac{1}{4}$) LM plateau progressively lengthens and the associated $T_{K}$ correspondingly diminishes, vanishing as the transition is approached from the Fermi liquid side (\fref{fig:fig4}).

  The behavior on the other side of the transition $\etiltwo < \tilde{\epsilon}_{2c}$ (cases (f-h) in \fref{fig:fig2}) is qualitatively distinct. Here the $T=0$ entropy is in all cases $\ln 2$ (with $T\Ximp = \tfrac{1}{4}$), characteristic of an unquenched doublet ground state. The stable FP is the spin-1/2 LM FP -- or 
equivalently the USC spin-1 FP~\cite{nozblan}, there being no distinction between them as FPs \emph{per se}.

  The QPT itself is of Kosterlitz-Thouless (KT) type. This is evident for example from NRG flows,
which indicate no separate, unstable critical FP, distinct from one of the stable FPs mentioned 
above. It is also evident in the behavior of the scale $T_{K}$, which as shown in \fref{fig:fig4} (inset) 
vanishes exponentially in $|\etiltwo -\tilde{\epsilon}_{2c}|^{-1}$ (rather than as a power-law) as the QPT is
approached from the Femi liquid side; and by the absence of a low-energy scale in the USC phase which vanishes as the transition is approached from that side. We add that the latter does not of course imply the inherent absence of a low-energy scale in the USC phase. For deep inside this phase (where $\nimp \simeq 2$) the effective low-energy model is spin-1 Kondo, as evident \eg in case (h) of  \fref{fig:fig3} from both the emergence of a near free spin-1 susceptibility with decreasing $T$ ($\Ximp \sim S(S+1)/3T$ with $S=1$), and from the intermediate $\Simp =\ln 3$ plateau indicative of a spin-1 local moment FP; reached before the crossover to the stable USC FP with $\Simp =\ln 2$, and from which  a  characteristic spin-1 Kondo scale $T_{K}^{S=1}$ may be identified (in parallel to that above for the Fermi liquid Kondo scale $T_{K}$). But this scale plays no role in the QPT \emph{per se}, and in contrast to 
the approach from the Fermi liquid phase, there is no vanishing scale on approaching the QPT from 
the USC side.

The behavior outlined above is not confined to the example illustrated: all continuous transitions 
are found to be of KT form. This is in fact to be expected. Hofstetter and Schoeller~\cite{HofSch} have consider the model (with $\up =U$) in the regime where the dot is doubly occupied by electrons, \ie $\nimp \simeq 2$ -- where from \eref{eq:nimpsymm}, $\nimp =2$ arises by symmetry along the line $y=-x$ in the  phase-plane (or close enough to it, in practice). A KT transition is likewise found in this case~\cite{HofSch}, and by continuity one would thus expect the same behavior to arise generally in the $(x,y)$-plane.

\begin{figure}
\includegraphics{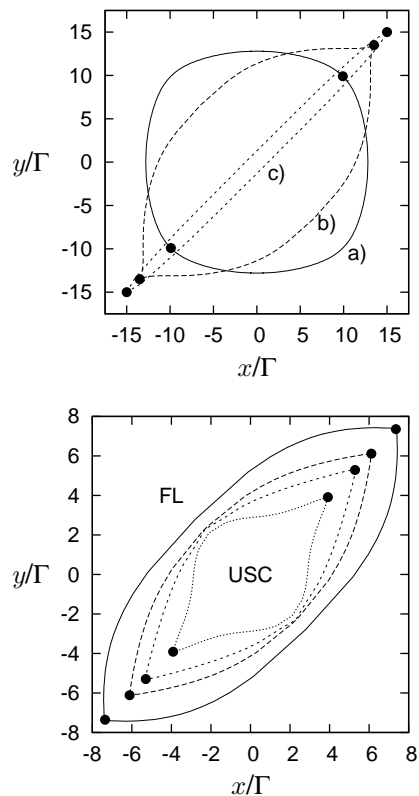}
\caption{\label{fig:fig5} Phase boundaries in the $(x,y)$-plane, separating the Fermi liquid phase (outer)
from USC spin-1 (inner). \emph{Top}: For $\util = 20$ and $\tilde{J}_{\mathrm{H}} =5$, varying $\uptil =0$ (a), $7.5$ (b) and $20$ (c). First-order transitions on the line $y=x$ are indicated by dots. ~\emph{Bottom}: For $\util =10$ and $\uptil = 5.25$, varying the exchange coupling $\tilde{J}_{\mathrm{H}} =5, 1, 0$ and $-0.5$ (outside to
inside respectively). Note the continued persistence of the USC \\spin-1 phase, even for weak antiferromagnetic exchange (see text).
}
\end{figure}

  We also note that the transition itself occurs generically in a mixed-valent regime of non-integral $\nimp$;
see \eg \fref{fig:fig3} (top, inset) where $\nimp$ varies continuously as the transition is crossed,
with $\nimp \simeq 1.4$ at the transition itself. This in turn means that even in a strongly correlated
regime it is not in general possible to construct, via a Schrieffer-Wolff~\cite{schriefferwolff} (SW) transformation from the original Anderson-like model, an effective low-energy \emph{spin} model applicable in the vicinity of the QPT. An exception to this is the vicinity of the line $y=-x$  along which, as above, $\nimp =2$ is guaranteed by symmetry. In this case, as shown in [\onlinecite{HofSch}], a SW transformation retaining solely the 2-electron $(1,1)$ triplet and $(2,0)$ singlet states of the isolated dot yields an effective two-spin, spin-$1/2$ Kondo model known~\cite{vojtabullahof02} to exhibit a KT transition.

  Phase diagrams obtained via NRG are shown in \fref{fig:fig5}. The top panel shows the effect of
varying the inter-level interaction $\up$ for fixed $U$ and $\J$, with behavior that parallels expectations 
from the isolated dot limit (\sref{sec:phaseoverview} and  \fref{fig:fig2}). The bottom panel
by contrast shows the effect of varying the exchange coupling $\J$ for fixed $U,\up$, including $\Jtil =0$
and an antiferromagnetic (AF) $\Jtil = -0.5$. Note that even for weakly AF exchange the USC spin-1 phase still persists, as considered further in \sref{sec:AFJ}, reflecting a ferromagnetic effective (RKKY) spin-spin interaction induced on coupling to the lead.


\subsection{First order transitions on $y=x$ line} 
\label{sec:firstorderthermo}

  We turn now to the first order transitions permitted by symmetry (\srefs{sec:symmetries}{sec:phaseoverview}) on the line $y=x$ ($\etiltwo =\etilone$ $\equiv \tilde{\epsilon}=\epsilon/\Gamma$).

\begin{figure}
\includegraphics{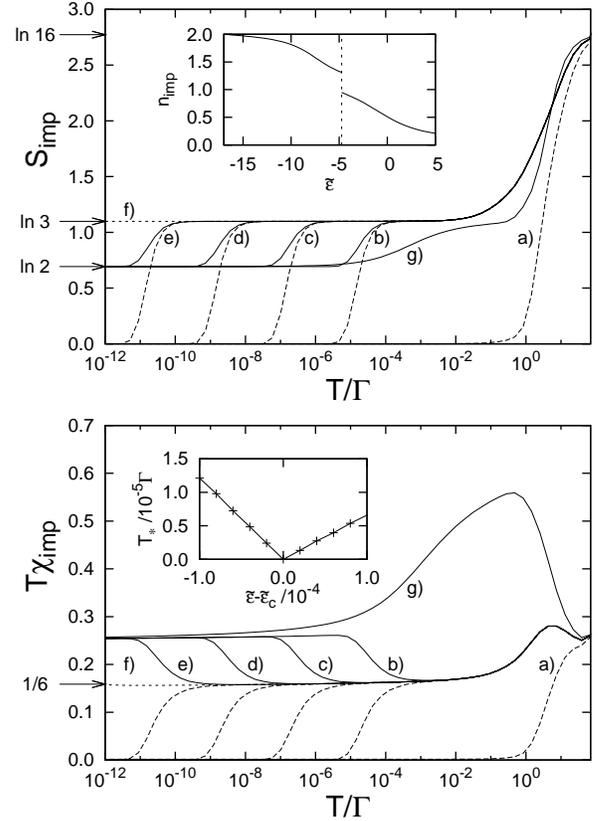}
\caption{\label{fig:fig6} $T$-dependence of the entropy $\simp$ (top panel) and spin susceptibility
$T\ximp$ (bottom) as the first-order transition on the line $\tilde{\epsilon}_{2}=\tilde{\epsilon}_{1}  \equiv \tilde{\epsilon}$ is approached and crossed from both sides of the transition; for fixed $\util = 20$, $\uptil =7$ and $\tilde{J}_{\mathrm{H}} =2$, the transition here occurring at $\tilde{\epsilon}_{c} = -4.73864856302929$.
Solid lines refer to the USC phase ($\tilde{\epsilon} < \tilde{\epsilon}_{c}$), dashed lines to the Fermi liquid (FL) phase. Shown for (a) $\tilde{\epsilon} =+5$ deep in the FL phase; (b-e) $\tilde{\epsilon} =\tilde{\epsilon}_{c} \pm 10^{-n}$ with $n=4,6,8,10$ respectively; (f) $\tilde{\epsilon}=\tilde{\epsilon}_{c}$; as well
as for (g) $\tilde{\epsilon} = -17$ at the $p$-$h$ symmetric point deep in  the USC phase.
The transition FP has a characteristic $S_{\mathrm{imp}} =\ln 3$ and $T\chi_{\mathrm{imp}} = \tfrac{1}{6}$,
as indicated; and persists down to $T=0$ precisely at the transition ((f)).
\emph{Inset, top panel}: $T=0$ impurity charge $\nimp$ \emph{vs} $\tilde{\epsilon}_{1} \equiv \tilde{\epsilon}$.
It changes discontinuously as the transition is crossed (dashed vertical line), from $\nimp \simeq 0.95$ to $1.31$.
\emph{Inset, bottom panel}: The low-energy scale $T_{*}$ (see text) vanishes linearly in 
$(\tilde{\epsilon}-\tilde{\epsilon}_{c})$ as the transition is approached, symptomatic of the level-crossing
nature of the transition.}
\end{figure}

To illustrate this, \fref{fig:fig6} shows the $T$-dependence of $\simp$ and $T\ximp$, again for fixed
$\util =20$, $\uptil = 7$ and $\Jtil =2$ (\emph{cf} \fref{fig:fig3}), as $\tilde{\epsilon}$ is varied and the transition is approached from both sides: $\tilde{\epsilon}=\tilde{\epsilon}_{c} \pm 10^{-n}$ with $n=4,6,8,10$ ( (b)-(e) respectively), as well as $\tilde{\epsilon}=\tilde{\epsilon}_{c} \simeq -4.7$ itself ((f)). Shown for comparison are the cases ((a)) $\tilde{\epsilon} = +5$ deep in the Fermi liquid regime, with $\tilde{\epsilon}$ here sufficiently large that the degenerate levels are barely occupied ($\nimp \simeq 0.25$, see top inset);
and $\tilde{\epsilon}=-17$ at the $p$-$h$ symmetric point deep in the USC phase ((g)).

The stable low-temperature FPs remain of course as before, \viz the FI FP for the Fermi liquid phase
$\tilde{\epsilon}>\tilde{\epsilon}_{c}$ where the global ground state is a singlet; and the USC FP
for $\tilde{\epsilon}<\tilde{\epsilon}_{c}$, with a doublet ground state. Close to the transition however
-- where the energy separation between these states is tending to zero (we denote its magnitude by $T_{*}$) 
-- the singlet and doublet states will appear effectively degenerate for temperatures $T\gtrsim T_{*}$; giving rise in consequence to an entropy plateau of $\Simp =\ln 3$ seen clearly in \fref{fig:fig6}, with a corresponding plateau
in the magnetic susceptibility of $T\Ximp =\tfrac{1}{6}$ (readily understood as the mean
$\langle(S^{z})^{2}\rangle \equiv \tfrac{1}{3}(1\times 0 + 2\times\tfrac{1}{4})$ for the quasidegenerate states).  These are signatures of the `transition fixed point' (TFP), characteristic of systems exhibiting a level-crossing
transition (see \eg [\onlinecite{TQD2009}]). On further reducing $T$ below $\sim T_{*}$, the system is seen to cross over from the TFP to one or other of the stable FPs (which crossover in effect defines $T_{*}$~\cite{footnoteone}).
Moreover, as the transition is approached, the low-energy scale $T_{*}$ vanishes -- linearly in $(\tilde{\epsilon}-\tilde{\epsilon}_{c})$ as shown in \fref{fig:fig6} (bottom inset), reflecting the level crossing character of the QPT. And since $T_{*}=0$ precisely at the transition, the TFP naturally persists down to $T=0$~\cite{TQD2009} (where the ground state consists of precisely degenerate global singlet and doublet states), as evident in case (f) of \fref{fig:fig6}.

 The behavior of the ($T=0$) excess impurity charge $\nimp$ is also shown in \fref{fig:fig6} (top inset). In contrast to the continuous KT transitions, $\nimp$ is seen to change discontinuously as the transition is
crossed, commensurate with the first order nature of the level-crossing transition. 

A partial progenitor of the latter behavior is in fact apparent in the trivial non-interacting limit, 
$U=0=\up =\J$. Taking even ($e$) and odd ($o$) combinations of the dot levels $1$ and $2$, \viz $d_{e\sigma}=(d_{1\sigma}+d_{2\sigma})/\sqrt{2}$ and $d_{o\sigma}=(d_{1\sigma}-d_{2\sigma})/\sqrt{2}$ (\emph{cf} \eref{eq:evenorb} with $\Gamma_{ii}\equiv \Gamma$, \eref{eq:equalgammas}), only the $e$-orbital tunnel couples directly to the lead and the non-interacting Hamiltonian $\hat{H}^{0}$ reduces to
\begin{equation}
\label{eq:NI}
\begin{split}
\hat{H}^{0}=&\tfrac{1}{2}(\epsilon_{1}+\epsilon_{2})(\hat{n}_{e}+\hat{n}_{o})+\sum_{\K ,\sigma}\sqrt{2}V(\ccre d_{e\sigma}^{\pd} + \mathrm{h.c.}) \\
&+\sum_{\sigma}\tfrac{1}{2}(\epsilon_{1}-\epsilon_{2})(d_{e\sigma}^{\dagger}d_{o\sigma}^{\pd} +\mathrm{h.c.})+\Hl
\end{split}
\end{equation}
(with $\Hl$ the lead Hamiltonian \eref{eq:hl}). In general, the $e$ and $o$ orbitals are coupled by the
penultimate term in \eref{eq:NI}. But for the case $\epsilon_{2}=\epsilon_{1} \equiv \epsilon$ of present
interest the Hamiltonian is separable,
$\hat{H}^{0} =\hat{H}^{0}_{e}+\hat{H}^{0}_{o}$, with $\hat{H}^{0}_{o} = \epsilon \hat{n}_{o}$ 
a free orbital with energy $\epsilon$. The transition in this case thus occurs trivially for $\epsilon =0$ (the $p$-$h$ symmetric point in the non-interacting limit) as the $o$-orbital -- which is unoccupied for 
$\epsilon >0$ -- moves across the Fermi level,  becoming singly occupied precisely at $\epsilon =0$ and doubly occupied for all $\epsilon <0$; such that $\nimp$ changes discontinuously from $1$ to $3$ as $\epsilon =0$ is crossed.

 With interactions present the situation is of course much less simple. For although the $o$-orbital
remains uncoupled from the lead when $\epsilon_{2}=\epsilon_{1}$, it is then coupled
to the $e$-orbital via the non-trivial interaction terms in $\Hdot$ (\eref{eq:hdot}), which acquire a 
rather complex (and physically unenlightening) form when expressed in terms of $e$ and $o$ operators.
We will return again to this case in \sref{sec:y=xdynamics}, from the perspective of dynamics and single-particle `renormalized levels'.


\subsection{Weakly antiferromagnetic $\mathbf{J}_{\mathbf{H}}^{}$} 
\label{sec:AFJ}

As noted above (\fref{fig:fig5}), the USC spin-1 phase survives even for weak antiferromagnetic (AF)
coupling $\J <0$, reflecting an effective ferromagnetic RKKY interaction induced on coupling to the lead.
For large enough AF $|\J|$ however, the situation is clearly different.
Only the \emph{singlet} state of the dot in the $(1,1)$ sector is relevant here, and on coupling to the leads one expects a global singlet ground state with a stable FI FP which is continuously connected to that of the normal FL
phases: no phase transitions then arise, and the USC phase is eliminated.

\begin{figure}
\includegraphics{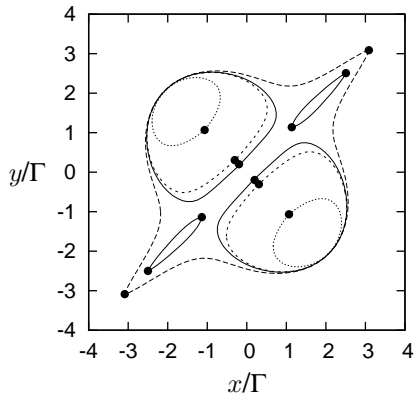}
\caption{\label{fig:fig7} Evolution of phase boundaries in the ($x,y$)-plane for antiferromagnetic
$\tilde{J}_{\mathrm{H}}<0$. Shown for fixed $\util = 10$ and $\uptil = 5.25$ (\emph{cf} \fref{fig:fig5} bottom),
with: $\tilde{J}_{\mathrm{H}} = -0.55$ (long dash line), $-0.558$ (solid), $-0.56$ (short dash) and
$-0.6$ (dotted). Interior regions in each case are the USC phase, exterior regions the Fermi liquid.
First-order transitions are indicated by dots, and occur on the lines $y=x$ \emph{and} $y=-x$.
}
\end{figure}

So how is the USC spin-1 phase destroyed as the strength of the AF coupling is
progressively increased? This is illustrated in \fref{fig:fig7}, showing phase boundaries in the $(x,y)$-plane
for fixed $\util =10$ and $\uptil = 5.25$ (as in \fref{fig:fig5}) for AF $\Jtil =-0.55, -0.558, -0.56$ and 
$-0.6$. For $\Jtil =-0.55$, the phase boundary has the same form as in \fref{fig:fig5}, consisting of the USC phase centred on ($x,y$)$=$($0,0$), separated from the exterior Fermi liquid phase by a single boundary line of 
KT transitions except on the line $y=x$ where a first order QPT arises.

On decreasing $\Jtil$ slightly to $-0.558$ however, the USC phase is seen to split into four distinct
domains -- symmetric as expected under both inversion, and reflection about $y=x$ -- with the $p$-$h$ symmetric point in particular now being in the FL phase. With a further slight decrease to $\Jtil = -0.56$, the two USC domains straddling $y=x$ are now eliminated, leaving two USC regions straddling the line $y=-x$. This behavior persists on further decreasing $\Jtil$, the remaining USC domains diminishing in extent until by $\Jtil \sim -1$ they too evaporate and the USC phase is eliminated entirely.

  Strikingly, as indicated in \fref{fig:fig7}, one also sees that as the USC phase fractionates, 
first order level-crossing transitions arise not only along $y=x$ (as expected on general
grounds and discussed in \sref{sec:firstorderthermo}); but also along the line $y=-x$.

To gain some insight into the above, note that the difference in energy (under $\Hdot$) between the  ($1,1$) singlet and triplet states of the isolated dot is $|E_{S}-E_{T}| =|\J|$. 
So for $|\J| = -\J \lesssim {\cal{O}}(\Gamma)$, and at least close enough to $p$-$h$ symmetry 
($x,y$)$=$($0,0$) (where $\nimp =2$), one expects it necessary to include both the ($1,1$) triplet \emph{and} 
singlet states in the low-energy dot manifold (higher dot states, such as ($2,0$), lie considerably higher in energy provided $\up$ is not within ${\cal{O}}(\J)$ of $U$). An effective low-energy model within this subspace may then be constructed via a Schrieffer-Wolff transformation~\cite{schriefferwolff}, the appropriate local unity operator 
being $\hat{1}=\hat{1}_{\mathrm{T}}+\hat{1}_{\mathrm{S}}$ with $\hat{1}_{\mathrm{T}}=\sum_{S^{z}} ~|S=1, S^{z}\rangle\langle S=1, S^{z}|$ and $\hat{1}_{\mathrm{S}}=|0,0\rangle\langle 0,0|$ (with $|S,S^{z}\rangle$ referring to the ($1,1$) triplet or singlet dot states). As discussed in Appendix A, the resultant effective model
is 
\begin{equation}
\label{eq:heff}
\hat{H}_{\mathrm{eff}}~=~ J_{1}~\hat{\mathbf{s}}_{1}\cdot\hat{\mathbf{s}}_{0}~+~
J_{2}~\hat{\mathbf{s}}_{2}\cdot\hat{\mathbf{s}}_{0}~-~I~\hat{\mathbf{s}}_{1}\cdot\hat{\mathbf{s}}_{2}~+~\Hl
\end{equation}
where as usual $\hat{\mathbf{s}}_{1}$ and $\hat{\mathbf{s}}_{2}$ are the spin-1/2 operators for levels 1 and 2, 
and $\hat{\mathbf{s}}_{0} =\sum_{\sigma ,\sigma^{\prime}}~f_{0\sigma}^{\dagger} 
\bm{\sigma}_{\sigma \sigma^{\prime}}^{\pd} f_{0\sigma^{\prime}}^{\pd}$ is the spin density of the conduction channel at the dot (with $f_{0\sigma}^{\dagger} = \tfrac{1}{\sqrt{N}} \sum_{\K} \ccre$ the creation operator
for the `$0$'-orbital of the Wilson chain~\cite{kww1,kww2} and $N$ the number of $\K$-states in
the lead). Only exchange scattering contributions to \eref{eq:heff} are shown explicitly, potential
scattering being omitted for clarity. The effective exchange couplings -- \viz the $J_{i}>0$
coupling spin $i=1$ or $2$ to the lead, and the direct spin exchange $I$ -- are naturally functions
of $x$ and $y$; expressions for them are given in Appendix A.

\Eref{eq:heff} is a two-spin Kondo model of the form studied in 
[\onlinecite{vojtabullahof02}], so its physics is understood~\cite{vojtabullahof02,HofSch}. A QPT,
occurring at a critical $I_{c}$, is obviously driven by the direct exchange $I$: for ferromagnetic
$\Delta I =I-I_{c} >0$, spins $1$ and $2$ form a spin-1 which is underscreened on coupling to the lead, resulting in a residual free spin-$\half$; while for AF $\Delta I =I-I_{c} <0$ by contrast, the local singlet Fermi liquid phase naturally arises. The resultant QPT is in general of KT form, with one pertinent exception~\cite{vojtabullahof02}: if $J_{2}=J_{1}$ in \eref{eq:heff}, then the Hamiltonian is separable into distinct singlet and triplet sectors for the spin $\hat{\mathbf{S}}=\hat{\mathbf{s}}_{1}+\hat{\mathbf{s}}_{2}$, specifically
\begin{equation}
\label{eq:heffseparable}
\hat{H}_{\mathrm{eff}}-\Hl~=~\hat{1}_{\mathrm{T}}\left( J_{1}\hat{\mathbf{S}}\cdot \hat{\mathbf{s}}_{0} 
-\tfrac{1}{4} I \right)\hat{1}_{\mathrm{T}}~+~\tfrac{3}{4} I~\hat{1}_{\mathrm{S}} ~~~~~~: J_{2}=J_{1}
\end{equation}
(on projecting \eref{eq:heff} with $\hat{1}=\hat{1}_{\mathrm{T}}+\hat{1}_{\mathrm{S}}$, and using 
$\hat{\mathbf{s}}_{1}\cdot\hat{\mathbf{s}}_{2} \equiv \half (\hat{S}^{2}-\tfrac{3}{2})$
together with $\hat{\mathbf{S}}\hat{1}_{\mathrm{S}} \equiv 0$ for all components of  $\hat{\mathbf{S}}$).

The resultant separability of the Hilbert space for $J_{2}=J_{1}$ means of course that a first order 
level-crossing transition can occur in this case. As shown in Appendix A (\eref{eq:Jtwo=Jone}), this is precisely the situation arising for the present problem along (and only along) the lines $y=x$ \emph{and} $y=-x$; explaining thereby the level-crossing transitions found in \fref{fig:fig7}~\cite{footnotetwo}.


\section{Dynamics and transport}
\label{sec:dynamics}

We now consider dynamics and transport, focussing on the $T=0$ zero-bias conductance and associated
phase shift, $\delta$. A `Friedel-Luttinger sum rule' for $\delta$ is derived, applicable to both the FL 
and USC spin-1 phases (\sref{sec:FLsumrule}). A generalization of Luttinger's integral
theorem for the USC phase is deduced, and its significant implications for the zero-bias conductance
determined (\sref{sec:zbcfsr}).

\subsection{Single-particle propagators}
\label{sec:GFs}

We first summarise basic results for the local single-particle propagators, embodied in the $2\times 2$ matrix
$\bm{G}(\omega)$. Its elements $G_{ij}(\omega)$ are the retarded Green functions for the dot levels,
$i,j \in{\{1,2\}}$ (as in \sref{sec:model}), related to the corresponding non-interacting
propagators $\bm{G}^{0}(\omega)$ by the Dyson equation
\begin{equation}
\label{eq:dyson}
[\bm{G}(\omega)]^{-1}~=~[\bm{G}^{0}(\omega)]^{-1}-\bm{\Sigma}(\omega)
\end{equation}
where $\bm{\Sigma}(\omega)$ is the $2\times 2$ self-energy matrix (with elements
$\Sigma_{ij}(\omega)= \Sigma_{ij}^{R}(\omega)-\I\Sigma_{ij}^{I}(\omega)$).
Using  equation of motion methods~\cite{Hewsonbook,Zubarev}
the elements of $[\bm{G}^{0}(\omega)]^{-1}$ are given by
\begin{equation}
\label{eq:g0inverse}
\left(\left[\bm{G}^{0}(\omega)\right]^{-1}\right)_{ij} ~=~(\omega^{+}-\epsilon_{i})\delta_{ij}~-
\Gamma_{ij}(\omega)
\end{equation}
where $\omega^{+}=\omega +\I 0+$, and $\Gamma_{ij}(\omega)$ is the hybridization function
\begin{equation}
\label{eq:gammaij}
\Gamma_{ij}(\omega)~=~\sum_{\K} \frac{\mathrm{V}_{i}\mathrm{V}_{j}}{\omega^{+}-\epsilon_{\K}}~~~\equiv~
\Gamma_{ij}^{R}(\omega) -\I \Gamma_{ij}^{I}(\omega)
\end{equation}
such that $\Gamma_{12}^{2}(\omega)=\Gamma_{11}(\omega)\Gamma_{22}(\omega)$ 
(for generality we allow here for arbitrary $V_{2},V_{1}$). For the standard~\cite{Hewsonbook} flat-band conduction spectrum/lead considered (\srefs{sec:model}{sec:thermo}), the imaginary part of the hybridization function 
$\Gamma_{ij}^{I}(\omega) = \Gamma_{ij}$ (\erefs{eq:gammaii}{eq:gamma12}) for $|\omega| <D$ and zero otherwise;
and the corresponding real part $\Gamma_{ij}^{R}(\omega =0)=0$ at the Fermi level ($\omega =0$).

 From \erefs{eq:dyson}{eq:g0inverse} it follows that $\bm{G}(\omega)$ has precisely the same algebraic 
structure as $\bm{G}^{0}(\omega)$, but with $\Gamma_{ij}(\omega)$ replaced by $\tilde{\Gamma}_{ij}(\omega)$ 
defined by:
\begin{equation}
\label{eq:gammatilde}
\tilde{\Gamma}_{ij}(\omega)= \Gamma_{ij}(\omega)+\Sigma_{ij}(\omega)
\end{equation}
Using \eref{eq:g0inverse} the propagators $G_{ij}(\omega)$ thus follow as
\begin{subequations}
\label{eq:Gijs}
\begin{align}
\label{eq:G11}
G_{11}(\omega) ~&=~\left(\omega^{+}-\epsilon_{2}-\tilde{\Gamma}_{22}(\omega)\right)~\mathrm{det}\bm{G}(\omega) \\
\label{eq:G22}
G_{22}(\omega) ~&=~\left(\omega^{+}-\epsilon_{1}-\tilde{\Gamma}_{11}(\omega)\right)~\mathrm{det}\bm{G}(\omega) \\
\label{eq:G12}
G_{12}(\omega)~&=~\tilde{\Gamma}_{12}(\omega)~\mathrm{det}\bm{G}(\omega)~~~~~\left(=G_{21}(\omega)\right) 
\end{align}
\end{subequations}
with the determinant given explicitly by:
\begin{equation}
\label{eq:detG}
\begin{split}
\mathrm{det}\bm{G}&(\omega)~=~\\
&\left[\left(\omega^{+}-\epsilon_{1}-\tilde{\Gamma}_{11}(\omega)\right) 
\left(\omega^{+}-\epsilon_{2}-\tilde{\Gamma}_{22}(\omega)\right) 
-\tilde{\Gamma}_{12}^{2}(\omega)\right]^{-1}
\end{split}
\end{equation}
These equations enable the propagators and their spectral densities $D_{ij}(\omega) = -\tfrac{1}{\pi}\mathrm{Im}G_{ij}(\omega)$ to be determined;
with self-energies obtained in practice via a matrix generalization of the standard NRG
method~\cite{bullahewprus,NRGrmp}, as outlined in Appendix B
and discussed further in \sref{sec:spectra}.

  It is also convenient at this point to note an  expression for the ($T=0$) excess impurity charge
$\nimp$; defined as the difference in charge of the entire system with/without the dot,
and hence $n_{\mathrm{imp}} =2\tfrac{(-1)}{\pi} \mathrm{Im}\int^{0}_{-\infty}d\omega$ $[\sum_{\K}G_{\K\K}(\omega)+G_{11}(\omega)+G_{22}(\omega)-\sum{_\K}G_{\K\K}^{o}(\omega)]$
in terms of the level propagators $G_{ii}(\omega)$, the propagators $G_{\K\K}(\omega)$ for the lead $\K$-states,
and their counterparts in the absence of the dot, $G_{\K\K}^{o}(\omega)=[\omega^{+}-\epsilon_{\K}]^{-1}$.
Using equation of motion methods it is simple to show that
$G_{\K\K}(\omega)=G_{\K\K}^{o}(\omega)+G_{\K\K}^{o}(\omega)\sum_{i,j}V_{i}G_{ij}(\omega)V_{j}G_{\K\K}^{o}(\omega)$,
\ie (via \eref{eq:gammaij}) 
$\sum{_{\K}}[G_{\K\K}(\omega)-G_{\K\K}^{o}(\omega)]= -\sum_{i,j}(\partial \Gamma_{ij}(\omega)/\partial\omega)G_{ij}(\omega)$, 
and hence:
\begin{equation}
\label{eq:nimpprops}
n_{\mathrm{imp}} ~=~
2~\tfrac{(-1)}{\pi} \mathrm{Im}~\int^{0}_{-\infty}d\omega ~ \sum_{i,j}G_{ij}(\omega)\left[\delta_{ij}-\frac{\partial\Gamma_{ij}(\omega)}{\partial\omega}\right]
\end{equation}
For the commonly considered case~\cite{Hewsonbook} of an infinitely wide flat band/lead, where
$\partial\Gamma_{ij}(\omega)/\partial\omega =0$ for \emph{all} $\omega$, $\nimp$ reduces to
$\nimp =2\int^{0}_{-\infty}d\omega [D_{11}(\omega)+D_{22}(\omega)]$ $\equiv \langle \hat{n}_{1}+\hat{n}_{2}\rangle$ -- \ie to the charge on the dot. For a finite lead bandwidth $D$ (as considered here) $\nimp$ is in practice very close to the dot charge, although  does not coincide identically with it. \\

 We focus now on the $T=0$ zero-bias conductance, given from \eref{eq:zbc} by
\begin{equation}
\label{eq:zbceespec}
\frac{G_{c}(T=0)}{G_{0}}~=~\frac{2e^{2}}{h}~~~\pi (\Gamma_{11}+\Gamma_{22})D_{ee}(\omega =0)
\end{equation}
and determined by the Fermi level value of the $ee$-spectrum, $(\Gamma_{11}+\Gamma_{22})D_{ee}(\omega) = \sum_{ij}\Gamma_{ij}\tfrac{(-1)}{\pi}\mathrm{Im}G_{ij}(\omega)$ (\eref{eq:gee}); an explicit expression
for which can be obtained using \erefs{eq:Gijs}{eq:detG} and the $\omega =0$ behavior of the
$\tilde{\Gamma}_{ij}(\omega)$. For both the normal Fermi liquid phase \emph{and} the USC phase, 
the imaginary part of the self-energy vanishes at the Fermi level,
\begin{equation}
\label{eq:Imsigma=0}
\Sigma_{ij}^{I}(\omega =0)~=~0.
\end{equation}
For the normal FL phase this is of course wholly familiar~\cite{Hewsonbook}. For the USC phase, we have established it by direct NRG calculation of the $\Sigma^{I}_{ij}(\omega)$ (it is also consistent with purely elastic
scattering at the Fermi level for a singular Fermi liquid~\cite{mehta}). Given \eref{eq:Imsigma=0}, $\tilde{\Gamma}_{ij}(0)$ follows from \erefs{eq:gammaij}{eq:gammatilde} as
$\tilde{\Gamma}_{ij}(0)=-\I \Gamma_{ij}+\Sigma^{R}_{ij}(0)$. Using this in \erefs{eq:Gijs}{eq:detG},
and defining renormalized single-particle levels in the usual way~\cite{Hewsonbook} by
\begin{equation}
\label{eq:epsiloni*}
\epsilon_{i}^{*}~=~\epsilon_{i}~+~\Sigma_{ii}^{R}(0),
\end{equation}
a simple if tedious calculation gives
\begin{equation}
\pi \left(\Gamma_{11}+\Gamma_{22}\right)D_{ee}(0)~=
~\frac{1}{1+\left[\frac{\epsilon_{1}^{*}\epsilon_{2}^{*}-\left(\Sigma_{12}^{R}(0)\right)^{2}
}{\epsilon_{1}^{*}\Gamma_{22}^{\pd}+\epsilon_{2}^{*}\Gamma_{11}^{\pd}-2\Gamma_{12}^{\pd}\Sigma_{12}^{R}(0)}\right]^{2}}~;\nonumber 
\end{equation}
or equivalently 
\begin{equation}
\label{eq:sin2d}
\pi \left(\Gamma_{11}+\Gamma_{22}\right)D_{ee}(0)~=~\mathrm{sin}^{2}\delta
\end{equation}
with $\delta$ given explicitly by:
\begin{equation}
\label{eq:deltarenorm}
\delta ~=~ \mathrm{arctan}\left[\frac{\epsilon_{1}^{*}\Gamma_{22}^{\pd}+\epsilon_{2}^{*}\Gamma_{11}^{\pd}-2\Gamma_{12}^{\pd}\Sigma_{12}^{R}(0)}{\epsilon_{1}^{*}\epsilon_{2}^{*}-\left(\Sigma_{12}^{R}(0)\right)^{2}}\right]~.
\end{equation}


\subsection{Friedel-Luttinger sum rule}
\label{sec:FLsumrule}
The quantity $\delta$ appearing in \erefs{eq:sin2d}{eq:deltarenorm} is simply the static scattering phase
shift, given alternatively by
\begin{equation}
\label{eq:phaseshiftdef}
\delta~=~\mathrm{Im} \ln\left(\mathrm{det}\bm{G}(0)\right)~~~~ \equiv ~\mathrm{Im} \ln\left(\mathrm{det}\bm{G}(\omega)\right)\big|^{\omega =0}_{\omega =-\infty}
\end{equation}
(the equivalence of \erefs{eq:deltarenorm}{eq:phaseshiftdef} is readily shown using
\eref{eq:detG});
the right hand side of \eref{eq:phaseshiftdef} also uses $\mathrm{arg}[\mathrm{det}\bm{G}(\omega =-\infty)] =0$,
as follows from \eref{eq:detG} together with the fact that as $|\omega| \rightarrow \infty$,
$\Sigma^{I}(\omega)$ vanishes while the $\Sigma_{ij}^{R}(\omega)$ tend to constants (the Hartree contributions).

  We now point out a general result for the phase shift $\delta$. From \erefs{eq:detG}{eq:Gijs}, it follows that
\begin{equation}
\frac{\partial}{\partial\omega} \ln \left(\mathrm{det}\bm{G}(\omega)\right)~=~-\sum_{i,j}\left(\delta_{ij}-\frac{\partial}{\partial\omega}\tilde{\Gamma}_{ij}(\omega) \right) G_{ji}(\omega)~. \nonumber
\end{equation}
Integrating this from $\omega =-\infty$ to $0$ (and noting \eref{eq:gammatilde}) then gives directly from \eref{eq:phaseshiftdef} that
\begin{equation}
\label{eq:FSR}
\delta ~=~ \tfrac{\pi}{2}\nimp ~+~ I_{L}
\end{equation}
where $\nimp$ is given by \eref{eq:nimpprops}, and (with $\mathrm{Tr}$ denoting a trace)
\begin{equation}
\label{eq:ILutt}
I_{L}~=~\mathrm{Im}~\mathrm{Tr}\int^{0}_{-\infty}d\omega ~ \frac{\partial\bm{\Sigma}(\omega)}{\partial\omega} \bm{G}(\omega)~
\end{equation}
is the Luttinger integral~\cite{luttingerfs,agdbook} (which is dimensionless).

  We emphasise that \eref{eq:FSR} is entirely general: applicable to both the normal Fermi liquid and the
USC phases (indeed its derivation does not even require a knowledge of \eref{eq:Imsigma=0}). For the
particular case of the FL phase, Luttinger's theorem gives $I_{L}=0$~\cite{luttingerfs,agdbook}; $I_{L}$
vanishing order by order in perturbation theory about the non-interacting limit,  reflecting adiabatic continuity to 
the non-interacting limit. In this case \eref{eq:FSR} reduces to the Friedel sum
rule~\cite{Hewsonbook,Langreth1966}, $\delta =\tfrac{\pi}{2}\nimp$, relating the scattering phase shift to the excess (`displaced') charged induced on addition of the dot/impurity to the system
(and with $\delta \in [0,2\pi]$ for a 2-level dot, since $\nimp \in [0,4]$).
More generally, however, $\delta$ and $\nimp$ are related by \eref{eq:FSR}, which we refer to as a
Friedel-Luttinger sum rule.

  The Luttinger integral for  the normal Fermi liquid phase is an intrinsic characteristic of it; $I_{L}=0$ holding independently of the underlying bare model parameters, provided only the system is a FL~\cite{luttingerfs,agdbook}.
As such, the Luttinger integral is the hallmark of the phase in a rather deep sense.

 The USC spin-1 phase by contrast is a singular Fermi liquid~\cite{mehta}. There is no
reason here to suppose $I_{L}=0$; and indeed it can be shown that the USC phase is
not perturbatively connected to the non-interacting limit of the model. But the obvious
question arises: as for the FL, does an analogous situation arise for the USC phase whereby the Luttinger
integral has a characteristic value for that phase? 

We answer this question affirmatively, by direct numerical calculation (and in several distinct ways). Since the self-energies $\bm{\Sigma}(\omega)$ and Green functions $\bm{G}(\omega)$ are calculable from NRG, 
we can calculate $I_{L}$ directly (\eref{eq:ILutt}) as an $\omega$-integral.  
Alternatively, $\nimp$ may be obtained from thermodynamic calculation 
(as in \sref{sec:thermo}) and $\delta$ from calculation of the $ee$-spectrum at the Fermi level alone
(as in \eref{eq:sin2d}, or alternatively \eref{eq:deltarenorm}); their difference then
giving the Luttinger integral,
$I_{L}=\delta -\tfrac{\pi}{2}\nimp$ from \eref{eq:FSR}.
We have confirmed that the same answer emerges in either way (and for the FL phase that $I_{L}=0$
thereby results). Namely, for any region of the ($x,y$)-plane where the system is in the USC phase,
the magnitude of the Luttinger integral is a constant, specifically:
\begin{equation}
\label{eq:modIL}
|I_{L}|~=~\frac{\pi}{2} ~~~~~: \mathrm{USC}
\end{equation}
We have repeated the calculations for many different values of the bare interaction parameters
$\util$, $\uptil$ and $\Jtil$. The same result emerges; and while the numerics obviously cannot
amount to a proof, we are confident in the validity of \eref{eq:modIL}.

  Although the magnitude of $I_{L}$ is constant throughout the USC phase, its sign is not. This is a
natural consequence of symmetry. By considering the symmetries of the propagators 
$\bm{G}(\omega) \equiv \bm{G}(\omega;x,y)$ and self-energies $\bm{\Sigma}(\omega;x,y)$ under a
particle-hole transformation (\sref{sec:symmetries}), it can be shown that the Luttinger integral
$I_{L} \equiv I_{L}(x,y)$ is odd under inversion,
\begin{equation}
\label{eq:ILphsymm}
I_{L}(x,y)~=~-I_{L}(-x,-y).
\end{equation}
In addition, as appropriate to the case $V_{2}=V_{1}$, the symmetries of $\bm{G}$, $\bm{\Sigma}$ under the
1-2 transformation (\eref{eq:12t}) lead to the rather obvious invariance under reflection
about the line $y=x$, 
\begin{equation}
\label{eq:IL12symm}
I_{L}(x,y)~=~I_{L}(y,x)~.
\end{equation}
With $|I_{L}|=\tfrac{\pi}{2}$, \eref{eq:ILphsymm} implies the existence of at least one bounding curve, of form
$y_{b}=f(x)$ with $f(x)=-f(-x)$, across which $I_{L}$ changes sign from $+\tfrac{\pi}{2}$
to $-\tfrac{\pi}{2}$; while \eref{eq:IL12symm} for the case $V_{2}=V_{1}$ implies that bounding curve
to be the line $y=-x$. In practice (by direct calculation, as above) only one bounding line is found; and for
the case $V_{2}=V_{1}$ in particular we find:
\begin{subequations}
\label{eq:ILsymmoverall}
\begin{align}
I_{L}(x,y)~&=~+\frac{\pi}{2} ~~~~~~ :~y >-x 
\label{eq:ILsymmoveralla}\\
&=~-\frac{\pi}{2} ~~~~~~ :~y<-x
\label{eq:ILsymmoverallb}
\end{align}
\end{subequations}


\subsection{Zero-bias conductance}
\label{sec:zbcfsr}

As above, $|I_{L}|=\tfrac{\pi}{2}$ is ubiquitous throughout the USC phase, as $|I_{L}|=0$ is throughout
the normal FL phase. This has immediate consequences for the behavior of the $T=0$ zero-bias conductance, given from
\erefss{eq:zbceespec}{eq:sin2d}{eq:FSR} by
\begin{equation}
\frac{G_{c}(T=0)}{G_{0}}~=~\frac{2e^{2}}{h}~\mathrm{sin}^{2}
\left(\frac{\pi}{2}\nimp +I_{L}\right)~.
\end{equation}
Since $\mathrm{sin}^{2}(\tfrac{\pi}{2}[\nimp \pm 1])=\mathrm{cos}^{2}(\tfrac{\pi}{2}\nimp)$, it follows that
\begin{subequations}
\label{eq:zbctwophases}
\begin{align}
\frac{G_{c}(T=0)}{\frac{2e^{2}}{h}G_{0}}~&=~\mathrm{sin}^{2}\left(\frac{\pi}{2}\nimp\right)~~~~~~
~~~~~~~~~:~\mathrm{FL} 
\label{eq:zbcFLphase}\\
&=~\mathrm{cos}^{2}\left(\frac{\pi}{2}\nimp\right)~~~~~~~~~~~~~~~:~\mathrm{USC}
\label{eq:zbcUSCphase}
\end{align}
\end{subequations}
for the FL and USC phases respectively. But as found in \sref{sec:thermo} (see \eg \fref{fig:fig3}), 
$\nimp$ varies continuously on crossing the line of Kosterlitz-Thouless transitions from the FL to the USC phase.
Hence from \eref{eq:zbctwophases} it follows that the zero-bias conductance must jump discontinuously
on crossing the QPT; the discontinuity on crossing from the FL to the USC phase being $\mathrm{cos}(\pi\nimp)$, with its sign determined by the value of $\nimp$ at the QPT. 
From direct calculation of single-particle spectra we will verify explicitly in \sref{sec:spectra} (see \eg \fref{fig:fig11}) that \erefs{eq:zbcFLphase}{eq:zbcUSCphase} are satsified throughout the two phases. 
\Eref{eq:zbctwophases} is of course equally applicable to the first order level-crossing transitions (\sref{sec:firstorderthermo}), although in this case $\nimp$ itself changes discontinuously as the transition is crossed (\fref{fig:fig6} top inset).


\section{Single-particle dynamics}
\label{sec:spectra}

 We turn now to $\omega$-dependent single-particle dynamics, here focussing primarily on the
 $D_{ee}(\omega)$ spectrum at $T=0$. The self-energies $\bm{\Sigma}(\omega)$ are obtained from NRG via a
generalization of the basic method~\cite{bullahewprus,NRGrmp} to the case of multilevel impurities/dots, as outlined in Appendix B. $\bm{\Sigma}$ is thereby calculated from
\begin{equation}
\label{eq:FoverG}
\bm{\Sigma}(\omega)~=~\bm{F}(\omega)~\left[\bm{G}(\omega)\right]^{-1}
\end{equation}
where the $2\times 2$ matrix $\bm{F}(\omega)$ has elements
\begin{equation}
\label{eq:Fij}
F_{ij}(\omega)~=~\langle\langle ~[\dides ,\hat{H}_{I}^{\pd}];d^{\dagger}_{j\sigma}~\rangle\rangle
\end{equation}
(using conventional notation~\cite{Hewsonbook,Zubarev} for the $\omega$-dependent 
Fourier transform of a generic retarded correlation function
$\langle\langle \hat{A}(t_{1});\hat{B}(t_{2})\rangle\rangle$$=$$-\I\theta(t_{1}$$-$$t_{2})\langle \{\hat{A}(t_{1}),\hat{B}(t_{2})\}\rangle$);  and
where  $\hat{H}_{I}$ denotes the interacting part of the dot Hamiltonian, given explicitly
for the present problem by (\eref{eq:hdot}) 
$\hat{H}_{I}= U\sum_{i} \hat{n}_{i\uparrow}^{\pd}\hat{n}_{i\downarrow}^{\pd} +\up\hat{n}_{1}^{\pd}\hat{n}_{2}^{\pd}
-\J~\hat{\mathbf{s}}_{1}\cdot\hat{\mathbf{s}}_{2}$. 
Using the self-energies, the fully interacting propagators are then obtained from the Dyson equation \eref{eq:dyson}.
As for single-level problems, calculation of $\bm{G}(\omega)$ in this way is numerically stable and accurate,
and guarantees satisfaction of spectral sum rules,  $\int^{\infty}_{-\infty}d\omega D_{ii}(\omega) =1$~\cite{bullahewprus,NRGrmp} (interleaved NRG/`z-averaging'~\cite{ztrick} also being used for optimal calculational accuracy).

\begin{figure}
\includegraphics{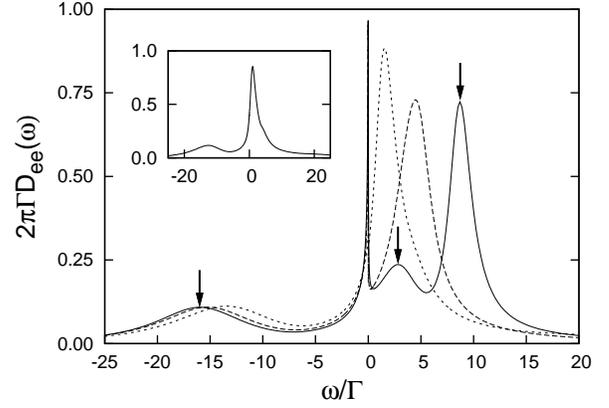}
\caption{\label{fig:fig8} 
Single-particle spectrum $2\pi\Gamma D_{ee}(\omega)$ \emph{vs} $\omega/\Gamma$ in the Fermi liquid
phase, for fixed level energy $\etilone =-\tfrac{1}{2}\util -\uptil$ (\ie $x=0$) on progressively
decreasing $\etiltwo$ towards the QPT (occurring at $\tilde{\epsilon}_{2c} = -6.536..$). 
For $\util =20$, $\uptil =7$ (\ie $\etilone = -17$) and $\tilde{J}_{\mathrm{H}} =2$, as in \fref{fig:fig3};
with $\etiltwo =$ +$1$ (solid line), $-3$ (long dash), $-6$ (short dash). 
Note that all three cases contain a narrow Kondo resonance straddling the Fermi level $\omega =0$. For vertical arrows, see text.~ 
\emph{Inset}: spectrum for $\tilde{\epsilon}_{2}=-6.6$ on just entering the USC phase, showing
the absence of a Kondo resonance.
}
\end{figure}

  As in \sref{sec:thermo} we consider explicitly $V_{2}=V_{1}$;
for which $\Gamma_{ij} \equiv \Gamma$ (\eref{eq:equalgammas}), with the $ee$-spectrum thus given 
(\eref{eq:gee}) by:
\begin{equation}
\label{eq:Deeequalgammas}
D_{ee}(\omega)~=~ \tfrac{1}{2}\left[ D_{11}(\omega)+D_{22}(\omega)+2D_{12}(\omega)\right]
\end{equation}

\Fref{fig:fig8} shows an `all scales' overview of $2\pi\Gamma D_{ee}(\omega)$ \emph{vs} $\tilde{\omega} \equiv \omega/\Gamma$, for fixed $\util =20$, $\uptil = 7$ and $\Jtil =2$ (as in \fref{fig:fig3} for thermodynamics), taking a vertical cut through the $(x,y)$-phase diagram: the energy of
level-1 is fixed at $\etilone = -\half\util -\uptil \equiv -17$ (\ie $x=0$), and $\etiltwo$ 
is progressively decreased through the Fermi liquid phase towards the transition ($\tilde{\epsilon}_{2c} = -6.536...$). 

The most important spectral feature is of course the clear Kondo resonance straddling the Fermi level. We consider it below, but first comment on the qualitative origin of the high-energy spectral features, evident most clearly in the three arrowed peaks shown in \fref{fig:fig8} for $\tilde{\epsilon}_{2}=+1$.
The corresponding evolution of $\nimp$ \emph{vs} $\tilde{\epsilon}_{2}$ is shown in
\fref{fig:fig3} (inset), from which it is seen than $\nimp \simeq 1.2$ for $\tilde{\epsilon}_{2}=+1$ --
sufficiently close to unity that we can interpret the high energy spectral features as removal or addition
excitations from the singly-occupied ($n_{1},n_{2}$)$=$($1,0$) state of the isolated dot. The removal 
excitation from dot level-$1$, contributing as such to the $D_{11}(\omega)$ constituent of $D_{ee}(\omega)$
(\eref{eq:Deeequalgammas}), thus
correponds trivially to $E_{D}(1,0)-E_{D}(0,0) =\epsilon_{1}$ (in the notation of {\sref{sec:phaseoverview}),
\ie lies below the Fermi level at ($\omega/\Gamma =$) $\tilde{\omega}=\tilde{\epsilon}_{1} \equiv -17$ here,
generating the lower `Hubbard satellite' seen clearly in \fref{fig:fig8}; its position, dependent at
this crude level of description only on $\tilde{\epsilon}_{1}$, varies only slightly on further decreasing $\tilde{\epsilon}_{2}$ in the FL regime.

  Two addition excitations lying above the Fermi level are also seen in \fref{fig:fig8} for
$\tilde{\epsilon}_{2}=+1$ (arrowed).  The lowest corresponds to electron addition to level-$1$, and hence shows up (again via $D_{11}(\omega)$) as an excitation at $E_{D}(2,0)-E_{D}(1,0) =\epsilon_{1}+U$; thus lying at $\tilde{\omega} = \tilde{\omega}_{+}= \tilde{\epsilon}_{1}+\util = +3$ as seen in the figure. The second excitation
corresponds to addition to level-$2$, contributes as such to the $D_{22}(\omega)$ constituent of $D_{ee}(\omega)$,
and thus corresponds to $E_{D}(1,1)-E_{D}(1,0)$. Since there are two distinct ($1,1$) dot states -- triplet and singlet -- two such excitations in principle arise; separated in energy by $\Jtil$ and occurring at 
$\tilde{\omega} = \tilde{\omega}_{T} = \tilde{\epsilon}_{2}+\uptil -\tfrac{1}{4}\Jtil$ (for  triplet ($1,1$)) and
$\tilde{\omega} = \tilde{\omega}_{S} = \tilde{\epsilon}_{2}+\uptil +\tfrac{3}{4}\Jtil$ for the singlet. As
evident from the figure, coupling to the leads in practice blurs these excitations, so that only a single spectral feature is seen.

On decreasing $\tilde{\epsilon}_{2}$ from $+1$ the $\tilde{\omega}_{S/T}$ excitations (which as above depend on $\tilde{\epsilon}_{2}$) decrease, and become comparable in energy to the $\tilde{\omega}_{+}$($=+3$) excitation; so that as seen in \fref{fig:fig8} the high-energy addition excitations in practice merge to a form a single peak, which on decreasing $\tilde{\epsilon}_{2}$ through the FL phase moves towards -- but does not reach -- the Fermi level. Instead, the single-particle spectrum in the immediate vicinity of the Fermi level $\omega =0$ is naturally
dominated by the narrow low-energy Kondo resonance evident in \fref{fig:fig8}.

\begin{figure}
\includegraphics{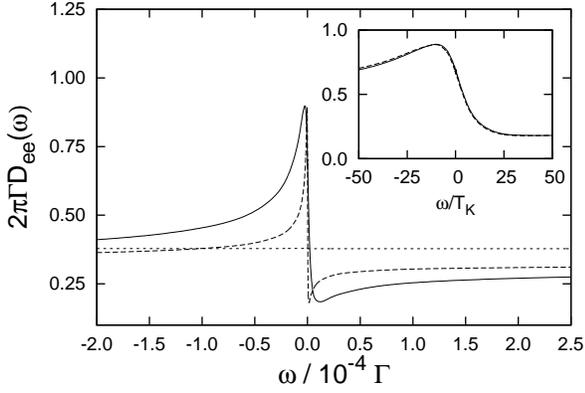}
\caption{\label{fig:fig9} For same parameters as \fref{fig:fig8}, showing a close-up
of the Kondo resonance on approaching the transition (at $\tilde{\epsilon}_{2c}=-6.536..$) from the FL side: $2\pi\Gamma D_{ee}(\omega)$ \emph{vs} $\omega/\Gamma$ for $\tilde{\epsilon}_{2}=-6.1$ (solid line) and
$-6.2$ (long dash). The Kondo resonance collapses `on the spot' as the QPT is approached and the Kondo scale $T_{K}\rightarrow 0$. The short dashed line shows the spectrum for $\tilde{\epsilon}_{2}=-6.54$ on just entering
the USC phase; it is featureless on these scales, with no Kondo resonance. ~\emph{Inset}: Scaling of
the Kondo resonance on approaching the QPT from the FL side. Both FL spectra in the main
figure collapse to a universal scaling resonance as a function of $\omega/T_{K}$ (their individual 
$T_{K}$s differ by more than an order of magnitude).
}
\end{figure}

  The evolution of the Kondo resonance itself is shown in close-up in \fref{fig:fig9}. As 
$\tilde{\epsilon}_{2} \rightarrow \tilde{\epsilon}_{2c}+$ from the FL side, it narrows
progressively -- reflecting the incipient vanishing of the Kondo scale $T_{K}$ known
from thermodynamics (\sref{sec:thermo}, \frefs{fig:fig3}{fig:fig4}) -- and collapses `on the spot'
at the transition itself, where $T_{K}$ vanishes. As a corollary,  in the USC phase just on the other side of the transition the Kondo resonance is simply absent; as seen in \fref{fig:fig9} (for $\tilde{\epsilon}_{2}=-6.54$)
where the USC spectrum is constant on the low $\tilde{\omega}=\omega/\Gamma$ scales shown. The inset to
\fref{fig:fig8} also shows this USC spectrum on an `all scales' level, showing that while the Kondo resonance is
absent here, the high-energy features discussed above evolve in a smooth way from those arising in
the FL phase.

  Since the Kondo scale $T_{K}$ vanishes as the QPT is approached from the FL side, one expects the 
Kondo resonance to exhibit universal scaling in terms of it.
That this is so is seen in \fref{fig:fig9} (inset), where both FL spectra shown in the main figure collapse to a universal scaling form as a function of $\omega/T_{K}$. Note also that while we have scaled the spectra here in terms of $T_{K}$ obtained from $\simp$ (as in \sref{sec:thermo}), we could equally have defined $T_{K}$ spectrally --
\eg via the width of the Kondo resonance -- and likewise obtained universal scaling behavior. The 
essential point is simply that there is only one vanishing low-energy scale as the QPT is approached, and 
different practical definitions of it are all fundamentally equivalent. 

\begin{figure}
\includegraphics{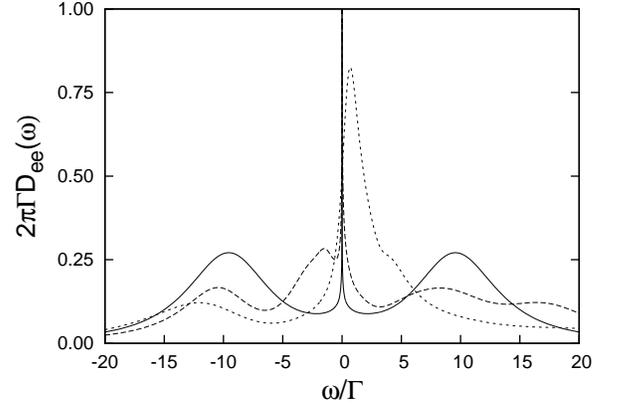}
\caption{\label{fig:fig10} USC phase. For same parameters as \frefs{fig:fig8}{fig:fig9},
$2\pi\Gamma D_{ee}(\omega)$ \emph{vs} $\omega/\Gamma$  for
$\tilde{\epsilon}_{2}=-7$ (short dash line), $-10$ (dashed) and the center of the USC phase $\tilde{\epsilon}_{2}=-17$ ($p$-$h$ symmetric point, solid line). The Kondo resonance which develops here is that for a spin-$1$ Kondo model ~\cite{KollerHewson}; see also \sref{sec:antires} (\fref{fig:fig13}).
}
\end{figure}

  The subsequent evolution of the spectrum in the USC phase is shown in \fref{fig:fig10}.
Not far into the USC phase ($\tilde{\epsilon}_{2}=-7$), the spectrum lacks a Kondo resonance, 
as above. However on further decreasing $\tilde{\epsilon}_{2}$, a second Kondo resonance straddling the Fermi level is seen to arise. It is in fact well developed already by $\tilde{\epsilon}_{2} =-10$, and narrows progressively as $\tilde{\epsilon}_{2}$ decreases towards the center of the USC phase at the $p$-$h$-symmetric point $\tilde{\epsilon}_{2}=-17$.
The origin of this behavior is readily guessed from $\nimp$ \emph{vs} $\tilde{\epsilon}_{2}$ 
(\fref{fig:fig3} inset). For although the transition itself corresponds to a `mixed valent' 
$\nimp \approx 1.4$, on entering the USC phase $\nimp$ increases quite rapidly; such that even by $\tilde{\epsilon}_{2}=-10$, $\nimp$ is close to $2$. Here one expects the system at low energies to be described asymptotically by a spin-1 Kondo model, and hence the second Kondo resonance to be of that ilk.
This is indeed so; we discuss it further in the context of \fref{fig:fig13} below.
High-energy spectral features in this regime are also naturally interpretable in terms of single-electron 
excitations to/from the ($1,1$) triplet ground state of the isolated dot; \eg at the $p$-$h$ symmetric point,
all addition/removal excitations to/from both levels $1$ and $2$ have the same magnitude,
$|\tilde{\epsilon}_{1}+\uptil -\tfrac{1}{4}\Jtil|$, giving rise to the symmetrically disposed Hubbard satellites at $|\tilde{\omega}| \simeq 10.5$ seen in \fref{fig:fig10}. 

  Finally and importantly, \fref{fig:fig11} verifies the predictions of \sref{sec:zbcfsr} 
for the behavior of the zero-bias conductance on crossing the QPT. The Fermi level spectrum
$2\pi\Gamma D_{ee}(0)$ \emph{vs} $\tilde{\epsilon}_{2}$ is shown in both phases, and compared
explicitly to  \eref{eq:zbctwophases} with $\nimp$ obtained from an independent
thermodynamic NRG calculation; the agreement being excellent.

\begin{figure}
\includegraphics{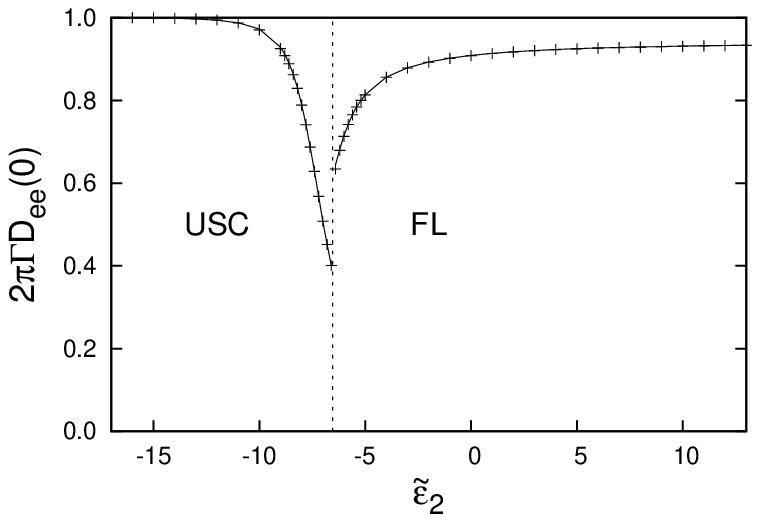}
\caption{\label{fig:fig11} 
$2\pi\Gamma D_{ee}(\omega =0)$ (equivalently the zero-bias conductance, see 
\erefs{eq:zbceespec}{eq:zbctwophases}) \emph{vs} $\tilde{\epsilon}_{2}$ in both phases on either side of the Kostelitz-Thouless transition (dashed vertical line at 
$\tilde{\epsilon}_{2c}=-6.536..$, same bare parameters as \frefseq{fig:fig8}{fig:fig10}). Crosses show
the $\omega =0$ spectra determined from NRG, while solid lines show $\mathrm{sin}^{2}(\tfrac{\pi}{2}\nimp)$ 
(in FL phase) and  $\mathrm{cos}^{2}(\tfrac{\pi}{2}\nimp)$ (in USC phase) with $\nimp$ obtained from a thermodynamic NRG calculation; verifying the predictions of \eref{eq:zbctwophases}. At the QPT,
$\nimp =1.43$, whence the spectrum/conductance decreases discontinuously on crossing from the FL to
the USC phase, $\nimp$ itself evolving continuously 
(\fref{fig:fig3} inset).
}
\end{figure}


\subsection{Kondo antiresonances}
\label{sec:antires}

 In the example considered above the QPT is associated with a collapsing Kondo
\emph{resonance} in the FL phase; and hence naturally with a \emph{decrease} in the zero-bias conductance 
on crossing into the USC phase. From \eref{eq:zbctwophases}, the latter behavior is generic
provided $\nimp$ at the transition lies in the interval $\nimp \in [\tfrac{1}{2},\tfrac{3}{2}]$ (by symmetry
\eref{eq:nimpsymm} we can consider $\nimp \in [0,2]$ rather than the full range $[0,4]$).
If however $\nimp$ at the QPT lies in the range $[\tfrac{3}{2},2]$ or $[0,\tfrac{1}{2}]$, then
\eref{eq:zbctwophases} predicts generically an \emph{increase} in the conductance on crossing from the FL
to the USC phase. One might thus intuitively expect such behavior to be associated with a vanishing Kondo
\emph{antiresonance} as the transition is approached from the FL side.

  That this indeed arises~\cite{HofSch} is illustrated in \fref{fig:fig12}, where dynamics on the line
$y=-x$ are considered; along which, by symmetry (\eref{eq:nimpsymm}),  $\nimp =2$ regardless of phase
(the spectra are likewise readily shown to be symmetric in $\omega$).
For bare interaction parameters $\util =20$, $\uptil =7$ and $\Jtil =2$, we decrease
($y/\Gamma \equiv$) $\tilde{y}=\tilde{\epsilon}_{2}+\tfrac{1}{2}\util+\uptil$ (\eref{eq:xandy}) across the
transition occuring at the critical $\tilde{y}_{c} = 6.36..$, from the FL side ($y>\tilde{y}_{c}$) to the USC phase.
\begin{figure}
\includegraphics{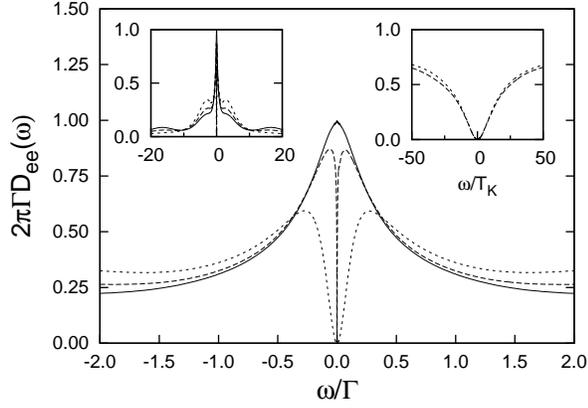}
\caption{\label{fig:fig12} 
$2\pi\Gamma D_{ee}(\omega)$ \emph{vs} $\omega/\Gamma$ along the line $y=-x$ (for same interaction parameters as \frefseq{fig:fig8}{fig:fig11}),
decreasing $\tilde{y}=\tilde{\epsilon}_{2}+\tfrac{1}{2}\util+\uptil$ through the QPT (at $\tilde{y}_{c} = 6.36..$)
from the FL side $y>\tilde{y}_{c}$. Shown for $\tilde{y}=6.55$ (short dash) and $6.40$ (long dash) in the
FL phase, and $\tilde{y}=6.30$ (solid) just into the USC phase.
A clear Kondo antiresonance at the Fermi level in the FL phase is seen (with $2\pi\Gamma D_{ee}(0)=0$). It collapses
`on the spot' as $T_{K} \rightarrow 0$ and the QPT is approached; and 
exhibits scaling as a function of $\omega/T_{K}$ as it does so (\emph{right inset}).
~\emph{Left inset}: as main figure, on an expanded $\omega/\Gamma$ scale. 
}
\end{figure}
As shown in the main figure $D_{ee}(\omega)$ indeed contains a Kondo antiresonance in the FL phase, 
here with $2\pi\Gamma D_{ee}(\omega =0)=0$ throughout. This antiresonance likewise vanishes on the spot as the transition is approached and the Kondo scale $T_{K} \rightarrow 0$;
and as it does so exhibits scaling as a function of $\omega/T_{K}$ (\fref{fig:fig12} right inset),
the low-frequency spectral behavior being $2\pi\Gamma D_{ee}(\omega) \propto (\omega/T_{K})^{2}$, symptomatic of a normal Fermi liquid.

  Note that the general predictions of \sref{sec:dynamics} are neatly exemplified by the above results:  
since $\nimp =2$ everywhere along the $y=-x$ line, \erefs{eq:zbceespec}{eq:zbctwophases} yield 
$2\pi\Gamma D_{ee}(0) = 1$ in the USC phase and $0$ in the FL phase (as confirmed by direct 
calculation, \frefs{fig:fig12}{fig:fig13}); and hence that the zero-bias conductance $G_{c}(T=0)/G_{0}$ 
in this case increases by precisely the conductance quantum  $2e^{2}/h$ on crossing the QPT 
into the USC phase.

  \Fref{fig:fig13} continues \fref{fig:fig12} into the USC phase, showing the
$ee$-spectra for $\tilde{y}=5, 3$ and $0$. As for its counterpart in \fref{fig:fig10}, the Kondo resonance 
which develops in the USC phase is that for a spin-$1$ Kondo model~\cite{KollerHewson}. As shown in Appendix A, its low-energy scale $T_{K}^{S=1}$ varies with the bare interaction parameters as (modulo an immaterial prefactor)
\begin{equation}
\label{eq:TKspin1ph}
T_{K}^{S=1} ~\propto ~ \exp \left( -\frac{\pi}{8} \left[\frac{(\util +\half \Jtil)^{2}-\tilde{y}^{2}}{\util +\half\Jtil} \right]\right)~.
\end{equation}
Hence on decreasing $\tilde{y}$ through the USC phase, the Kondo resonance becomes increasingly narrow as 
$T_{K}^{S=1}$ decreases towards its smallest (but non-zero) value occurring at the $p$-$h$ symmetric point
$\tilde{y}=0$ ($=\tilde{x}$); and as shown in \fref{fig:fig13}, universal spectral scaling as a function of
$\omega/T_{K}^{S=1}$ thereby arises. 

\Fref{fig:fig13} (inset) also shows the clear cusp-like behavior of the spin-$1$ Kondo resonance as $|\omega| \rightarrow 0$, known from study of the spin-$1$ Kondo model itself~\cite{KollerHewson} (with spectra inferred from the $t$-matrix of the Kondo model). This behavior is characteristic of the singular Fermi liquid~\cite{mehta} 
nature of the underscreened spin-$1$ phase; specifically the weak ferromagnetic coupling of the residual
spin-$\half$ to the metallic lead, resulting in logarithmic corrections to Fermi liquid behavior.
As $|\omega|/T_{K}^{S=1} \rightarrow 0$ we find
\begin{equation}
2\pi \Gamma D_{ee}(\omega =0) ~\sim ~ 1~-
~\frac{b}{\ln^{2}(|\omega|/T_{K}^{S=1})} 
\end{equation}
(with $b$ a constant), the leading logarithmic correction here stemming from the leading
low-$\omega$ behavior of the self-energies $\Sigma^{I}_{ij}(\omega) \sim 1/\ln^{2}(|\omega|/T_{K}^{S=1})$; and which form is in agreement with that of [\onlinecite{KollerHewson}] for the spin-$1$ Kondo model itself.

\begin{figure}
\includegraphics{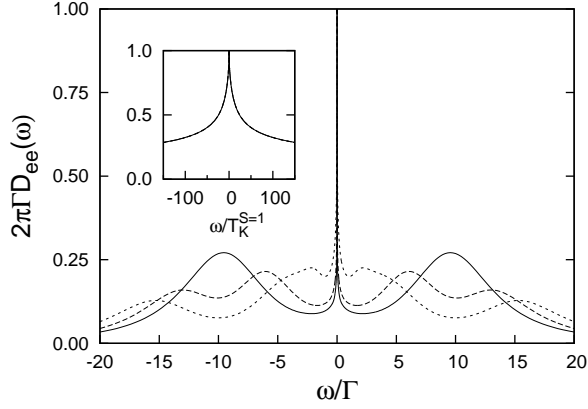}
\caption{\label{fig:fig13} Continuing \fref{fig:fig12} into the USC phase along the line $y=-x$: ~
$2\pi\Gamma D_{ee}(\omega)$ \emph{vs} $\omega/\Gamma$ for $\tilde{y}=5$ (short dash), $3$ (dash) and
the $p$-$h$ symmetric point $\tilde{y}=0 =\tilde{x}$ (solid).  The Kondo resonance developing in
the USC phase is that for the spin-$1$ Kondo model~\cite{KollerHewson}; and (\emph{inset}) the three spectra  in the main figure show low-energy universal scaling as a function of $\omega/T_{K}^{S=1}$ with $T_{K}^{S=1}$ the spin-$1$
Kondo scale (see text).
}
\end{figure}


\subsection{The $\mathbf{y=x}$ line}
\label{sec:y=xdynamics}

As considered in \sref{sec:firstorderthermo} in regard to thermodynamics, 
the transition occurring along the line $\epsilon_{2}=\epsilon_{1}$ (\ie $y=x$)
is a first order level-crossing transition, as permitted by symmetry for $V_{2}=V_{1}$.
Here we consider the $\epsilon_{2}=\epsilon_{1}$ line again, from the perspective of dynamics, and the
resultant channel separability arising in the `even/odd' representation as now discussed.


\subsubsection{Even/odd basis}
\label{sec:eobasis}

In previous sections the elements of the Green function matrix $\bm{G}(\omega)$ have been considered as
the propagators for the dot levels, \viz $G_{ij}(\omega)$ with $i,j \in \{1,2\}$. Equally, one can
take even/odd combinations of the dot levels, \viz
$d_{e\sigma}=(d_{1\sigma}+d_{2\sigma})/\sqrt{2}$ and
$d_{o\sigma}=(d_{1\sigma}-d_{2\sigma})/\sqrt{2}$, and consider $\bm{G}(\omega)$ in an $e/o$ representation; with
elements $G_{\alpha\beta}(\omega)$ given explicitly by
$G_{\substack{ee\\oo}}(\omega) = \tfrac{1}{2}[G_{11}(\omega)+G_{22}(\omega)\pm 2G_{12}(\omega)]$, with
$G_{eo}(\omega)=\tfrac{1}{2}[G_{11}(\omega)-G_{22}(\omega)]$ ($=G_{oe}(\omega)$) for the off-diagonal elements.
For $\epsilon_{2} \neq \epsilon_{1}$ in general, there is no particular advantage in working with the
$e/o$ representation. However along the line $\epsilon_{2}=\epsilon_{1}$ where levels $1$ and $2$ are equivalent by symmetry, $G_{11}(\omega)=G_{22}(\omega)$ and hence the off-diagonal $G_{eo}(\omega)=0$. $\bm{G}(\omega)$ in the $e/o$ representation is then purely diagonal for \emph{all} $\omega$, with elements:
\begin{equation}
\label{eq:Geeoo}
G_{\substack{ee\\oo}}(\omega) ~=~ G_{11}(\omega)~\pm ~G_{12}(\omega)
\end{equation}

Using \erefss{eq:Gijs}{eq:detG}{eq:gammatilde}, one obtains
\begin{subequations}
\begin{align}
G_{ee}(\omega)~&=~\left[\omega^{+}-\epsilon -2\Gamma(\omega) -\Sigma_{ee}(\omega)\right]^{-1}
\label{eq:Gee} \\
G_{oo}(\omega)~&=~\left[\omega^{+}-\epsilon -\Sigma_{oo}(\omega)\right]^{-1}
\label{eq:Goo}
\end{align}
\end{subequations}
where $\epsilon \equiv \epsilon_{1}=\epsilon_{2}$ denotes the common level energy,
the hybridization function is $\Gamma(\omega)$ ($ \equiv \Gamma_{ij}(\omega)$, \eref{eq:gammaij} with $V_{2}=V_{1}$), and the $ee/oo$ self-energies are given simply by:
\begin{equation}
\Sigma_{\substack{ee\\oo}}(\omega) ~=~ \Sigma_{11}(\omega)~\pm ~\Sigma_{12}(\omega)
\end{equation}
Notice from \eref{eq:Goo} that there is no direct hybridization ($\Gamma(\omega)$) contribution to $G_{oo}(\omega)$, reflecting the fact (\sref{sec:firstorderthermo}) that for $\epsilon_{2}=\epsilon_{1}$
the $o$-orbital is not directly coupled to the lead. In the non-interacting limit the $o$-level is 
thus entirely free, $G_{oo}^{0}(\omega)=[\omega^{+}-\epsilon]^{-1}$; but in general the $o/e$ levels
are coupled via interactions, as embodied in $\Sigma_{oo}(\omega)\neq 0$.

Since $\bm{G}(\omega)$ is diagonal in the $e/o$ representation,
$\mathrm{det}\bm{G}(\omega)=G_{ee}(\omega)G_{oo}(\omega)$, and hence from \eref{eq:phaseshiftdef}
the static phase shift $\delta$ is separable into $e$ and $o$ channels,
\begin{equation}
\delta ~=~\delta_{e}~+~\delta_{o}~.
\end{equation}
A short calculation using \eref{eq:phaseshiftdef} (together with $\Sigma_{\alpha\alpha}^{I}(\omega=0)=0$ from
\eref{eq:Imsigma=0}) then gives
\begin{subequations}
\begin{align}
\delta_{e}~&=~\mathrm{arctan}\left(\frac{2\Gamma}{\epsilon_{e}^{*}}\right) 
\label{eq:deltaeven}\\
\delta_{o}~&=~\mathrm{arctan}\left(\frac{0+}{\epsilon_{o}^{*}}\right)~\equiv ~ \pi~\theta(-\epsilon_{o}^{*})
\label{eq:deltaodd}
\end{align}
\end{subequations}
where each $\delta_{\alpha} \in [0,\pi]$, $\Gamma$ ($=\Gamma^{I}(\omega=0)$) is the usual hybridization
strength (\eref{eq:equalgammas}), and $\theta(u)$ is the unit step function. The $\epsilon_{\alpha}^{*}$ 
denote the renormalized $e/o$ levels, given by (\emph{cf} \eref{eq:epsiloni*})
\begin{equation}
\label{eq:renormeo}
\epsilon_{\alpha}^{*}~=~\epsilon + \Sigma_{\alpha\alpha}^{R}(0)
\end{equation}
with $\epsilon_{\alpha}^{*} \equiv \epsilon_{\alpha}^{*}(x)$ such that 
$\epsilon_{\alpha}^{*}(x)=-\epsilon_{\alpha}^{*}(-x)$ (via a $p$-$h$ transformation, \sref{sec:symmetries}).
Likewise, considering $\partial \ln G_{\alpha\alpha}(\omega)/\partial\omega$ and repeating the calculation
leading to \eref{eq:FSR}, gives
\begin{equation}
\label{eq:deltaalphalutt}
\delta_{\alpha}~=~\tfrac{\pi}{2}n_{\mathrm{imp},\alpha}~+~I_{L}^{\alpha}
\end{equation}
where (\emph{cf} \eref{eq:ILutt})
\begin{equation}
\label{eq:ILuttalpha}
I_{L}^{\alpha}~=~\mathrm{Im}~\int^{0}_{-\infty}d\omega ~ \frac{\partial\Sigma_{\alpha\alpha}(\omega)}{\partial\omega} G_{\alpha\alpha}(\omega)~
\end{equation}
is a Luttinger integral for channel $\alpha =e$ or $o$ (with
$I_{L}^{\alpha} \equiv I_{L}^{\alpha}(x)$ such that $I_{L}^{\alpha}(x)=-I_{L}^{\alpha}(-x)$
under inversion); and $n_{\mathrm{imp},\alpha}$ is the excess impurity charge associated
with channel $\alpha$, given by
\begin{subequations}
\label{eq:nimpalphas}
\begin{align}
n_{\mathrm{imp},e} ~&=~
2~\tfrac{(-1)}{\pi} \mathrm{Im}~\int^{0}_{-\infty}d\omega ~ G_{ee}(\omega)\left[1-2\frac{\partial\Gamma(\omega)}{\partial\omega}\right] \\
n_{\mathrm{imp},o} ~&=~
2~\tfrac{(-1)}{\pi} \mathrm{Im}~\int^{0}_{-\infty}d\omega ~ G_{oo}(\omega)
\end{align}
\end{subequations}
such that the overall $\nimp =n_{\mathrm{imp},e}+n_{\mathrm{imp},o}$ (\eref{eq:nimpprops}),
and with $n_{\mathrm{imp},\alpha}(x)=2-n_{\mathrm{imp},\alpha}(-x)$. Since the behavior of relevant
quantities under inversion $x \rightarrow -x$ is as specified above, we can focus on $x=\epsilon+\tfrac{1}{2}U+U^{\prime} \geq 0$; and do so in the following.


\subsubsection{Results}

\begin{figure}
\includegraphics{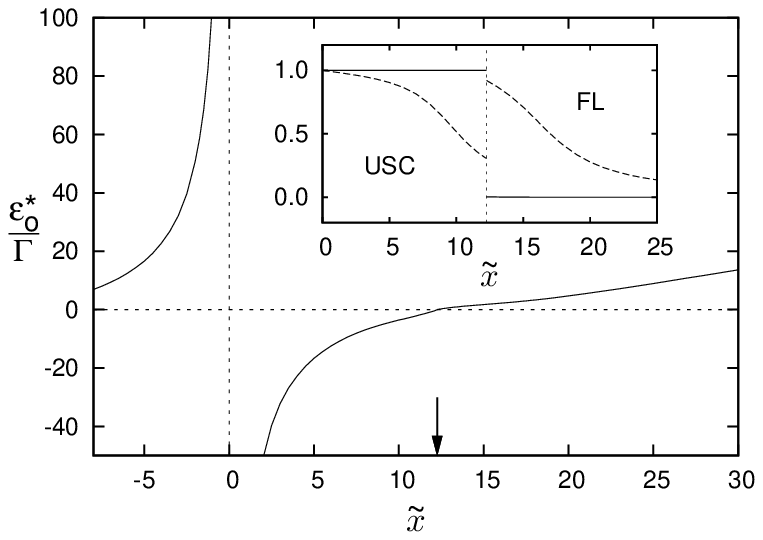}
\caption{\label{fig:fig14} 
Renormalized odd level $\epsilon_{o}^{*}/\Gamma$ (\erefs{eq:renormeo}{eq:epsilonostar}) \emph{vs} $\tilde{x}
=x/\Gamma =\tilde{\epsilon} +\tfrac{1}{2}\util +\uptil$ (for fixed $\util =20, \uptil =7$ and $\Jtil =2$).
The level-crossing transition at $\tilde{x}_{c}=12.26..$ is marked by an arrow; 
$\epsilon_{o}^{*} \sim x-x_{c} \equiv\epsilon -\epsilon_{c}$ vanishes linearly as the transition is approached. 
~\emph{Inset}: Behavior of $n_{\mathrm{imp},o}$
(solid line) and $n_{\mathrm{imp},e}$ (dashed) on crossing from the FL phase ($\tilde{x}>\tilde{x}_{c}$)
to the USC phase. Both jump discontinuously on crossing the transition $\mathrm{FL}\rightarrow \mathrm{USC}$, $n_{\mathrm{imp},o}$ upwards from $0$ to $1$ and $n_{\mathrm{imp},e}$ downwards as indicated; such that the overall
$\nimp = n_{\mathrm{imp},o}+n_{\mathrm{imp},e}$ increases as shown in \fref{fig:fig6} (top inset).
}
\end{figure}
The charges $n_{\mathrm{imp},\alpha}$ may be calculated directly from NRG. Their evolution with $\tilde{x}=x/\Gamma$ is illustrated in \fref{fig:fig14} (inset), on decreasing $\tilde{\epsilon} =\epsilon/\Gamma$
with $\util, \uptil$, $\Jtil$ fixed at the same values used in \fref{fig:fig6} for thermodynamics; the level-crossing transition here occurring at 
$\tilde{x}_{c}=12.26..$ ($\tilde{\epsilon}_{c}=-4.73..$). On crossing the transition from the FL side
($\tilde{x}>\tilde{x}_{c}$) to the USC phase, $n_{\mathrm{imp},o}$ increases discontinuously from $0$ -- 
found to be its value for \emph{all} $\tilde{x}>\tilde{x}_{c}$ -- to $n_{\mathrm{imp},o}=1$, which constant value is likewise found throughout the USC phase $\tilde{x}<\tilde{x}_{c}$. For the $e$-channel by contrast $n_{\mathrm{imp},e}$ is not fixed in either phase; but it too jumps discontinuously, \emph{decreasing} as the FL$\rightarrow$USC transition is crossed. The latter behavior is physically natural, since the piling of charge into the $o$-orbital which accompanies the transition increases Coulomb repulsions with electrons in the $e$-orbital, which the concomitant reduction in $n_{\mathrm{imp},e}$ acts to offset. 
The two effects do not however cancel, the overall
$\nimp = n_{\mathrm{imp},o}+n_{\mathrm{imp},e}$ (shown in \fref{fig:fig6}, top inset) increasing as the 
transition is crossed. The behavior just described is redolent of, but distinct from, that occurring in the non-interacting limit discussed in \sref{sec:firstorderthermo}; where at the transition, in that case occuring for $\tilde{\epsilon}_{c}=0$, $n_{\mathrm{imp},o}$ jumps discontinuously from $0$  to $2$ but with no concomitant change in $n_{\mathrm{imp},e}$ since there are no interactions present. Moreover since the transition is generically accompanied by occupancy of the $o$-orbital, one intuitively expects the requisite critical $\tilde{\epsilon}_{c}$ for the transition with interactions present to be reduced below its non-interacting counterpart $\tilde{\epsilon}_{c}=0$, in order to offset the increased interactions; as indeed is found. We also add that the behavior found is not specfic to the interaction parameters used for illustration; in particular that
\begin{subequations}
\label{eq:nimpdisc}
\begin{align}
n_{\mathrm{imp},o}~&=~ 0~~~~~~~~:~\mathrm{FL},~~~~ \tilde{x}>\tilde{x}_{c} \\
&=~1~~~~~~~~:~\mathrm{USC}, ~\tilde{x}<\tilde{x}_{c} 
\end{align}
\end{subequations}
is found to occur ubiquitously.

  We consider now the static renormalized levels, calculable from \eref{eq:renormeo};
or, for $\epsilon_{o}^{*}$, equivalently from
\begin{equation}
\label{eq:epsilonostar}
\epsilon_{o}^{*}~=~\frac{\epsilon}{1+F_{oo}^{R}(\omega =0)}
\end{equation}
where $F_{oo}(\omega) = F_{11}(\omega)- F_{12}(\omega)$  (with $F_{oo}^{R}(\omega)$ its real part),
and the $F_{ij}(\omega)$ are given by \eref{eq:Fij} and calculated directly via NRG
[\eref{eq:epsilonostar} follows from \eref{eq:renormeo} together with \eref{eq:B5}  
in the diagonal $e/o$ representation]. The generic $\tilde{x}$-dependence of $\tilde{\epsilon}_{o}^{*}=\epsilon_{o}^{*}/\Gamma$ is illustrated in \fref{fig:fig14}.
It evolves continuously for all $\tilde{x}>0$ (the divergence on approaching
the $p$-$h$ symmetric point $\tilde{x}=0$ at the center of the USC phase
 reflects via \eref{eq:epsilonostar} the fact that $F_{oo}^{R}(0) \rightarrow -1^{\mp}$ as 
$x \rightarrow 0^{\pm}$). In particular, in the USC phase $0<\tilde{x}<\tilde{x}_{c}$, 
the renormalized level $\epsilon_{o}^{*}<0$, while for $\tilde{x}>\tilde{x}_{c}$ in the FL phase, 
$\epsilon_{o}^{*}>0$; the level vanishing linearly as the QPT is crossed,
\begin{equation}
\label{eq:olevelvanish}
\epsilon_{o}^{*}~\overset{x\rightarrow x_{c}}\sim~ x-x_{c} ~ \equiv \epsilon -\epsilon_{c} ~.
\end{equation}
And for large enough $\tilde{\epsilon} \gg 1$, where both the $e$ and $o$ levels are in practice empty
and interaction effects embodied in $\Sigma_{oo}$ are thus irrelevant, $\epsilon_{o}^{*} \rightarrow \epsilon$ (the `bare' level energy, see \eref{eq:renormeo}).

  The above results then enable the $o$-channel Luttinger integral $I_{L}^{o}$ (\eref{eq:ILuttalpha}) to
be deduced. Since $\epsilon_{o}^{*} <0$ [$>0$] in the USC [FL] phase, \eref{eq:deltaodd} gives a phase shift $\delta_{o}=\pi$ in the USC phase $0<\tilde{x}<\tilde{x}_{c}$, and $\delta_{o} =0$ in the FL phase $\tilde{x}>\tilde{x}_{c}$. Combining this with \eref{eq:nimpdisc} for $n_{\mathrm{imp},o}$, the Luttinger integral 
$I_{L}^{o}=\delta_{o}-\tfrac{\pi}{2}n_{\mathrm{imp},o}$ (\eref{eq:deltaalphalutt}) follows directly as
\begin{subequations}
\label{eq:Iluttdisc}
\begin{align}
I_{L}^{o}~&=~ 0~~~~~~~~~:~\mathrm{FL},~~~~~~ \tilde{x}>\tilde{x}_{c} \\
&=~\frac{\pi}{2}~~~~~~~~:~\mathrm{USC}, ~~0<\tilde{x}<\tilde{x}_{c} 
\label{eq:IluttdiscUSC}
\end{align}
\end{subequations}
-- which result we have also verified by direct computation of $I_{L}^{o}$ itself, \eref{eq:ILuttalpha}.

For the $e$-channel by contrast, direct calculation of $I_{L}^{e}$ gives $I_{L}^{e}=0$ in both the FL phase 
\emph{and} the USC phase,
\begin{equation}
\label{eq:ILutteven}
I_{L}^{e}~=~0 ~~~~~~~:~ \mathrm{FL~and~USC}~.
\end{equation}
The total Luttinger integral $I_{L}=I_{L}^{o}+I_{L}^{e}$ thus vanishes as required~\cite{luttingerfs,agdbook}
throughout the FL phase; while for the USC phase \erefs{eq:IluttdiscUSC}{eq:ILutteven} agree as they must
with the general result \eref{eq:ILsymmoveralla} for $I_{L}(x,y)$ (which is not confined to the $y=x$ line).
Note further, using \eref{eq:ILutteven}, that \erefs{eq:deltaeven}{eq:deltaalphalutt} give
\begin{equation}
\label{eq:evenrenorm}
\epsilon_{e}^{*}~=~2\Gamma~\mathrm{tan}\left(\tfrac{\pi}{2}\left[1-n_{\mathrm{imp},e}\right]\right)~~~~
:~ \mathrm{FL~and~USC}
\end{equation}
independently of the phase (as again verified by separate calculation of $\epsilon_{e}^{*}$ 
and $n_{\mathrm{imp},e}$). From the $\tilde{x}$-dependence of $n_{\mathrm{imp},e}$ illustrated
in \fref{fig:fig14} (inset), \eref{eq:evenrenorm} shows that $\epsilon_{e}^{*}$ progressively decreases as $\tilde{x}$ is decreased through the FL phase, increases discontinuously as the FL$\rightarrow$USC transition is crossed, and in the USC phase decreases monotonically as $\tilde{x}$ is decreased towards the $p$-$h$ symmetric point $\tilde{x}=0$, where $n_{\mathrm{imp},e}=1$ and hence $\epsilon_{e}^{*}=0$ (and with $\epsilon_{e}^{*}(x)$
for $\tilde{x}<0$ following from the symmetry $\epsilon_{e}^{*}(x)=-\epsilon_{e}^{*}(-x)$). 
\begin{figure}
\includegraphics{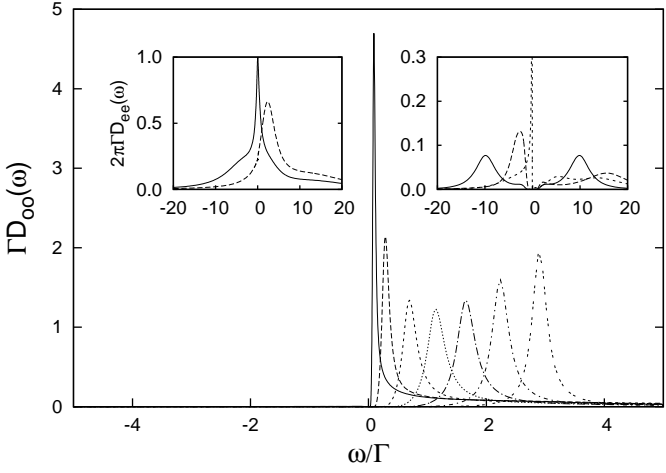}
\caption{\label{fig:fig15} 
$o$-orbital spectrum $\Gamma D_{oo}(\omega)$ \emph{vs} $\omega/\Gamma$ on decreasing $\tilde{x}=x/\Gamma$ in the FL phase (for same interaction parameters as \fref{fig:fig14}, with the critical $\tilde{x}_{c}=12.26..$). Shown for $\tilde{x}=18,17,16,15,14,13$ and $12.5$ (from right to left). 
~\emph{Right inset}: as main figure, but in the USC phase for $\tilde{x}=12$ (dotted line), $7$ (dashed) and at
the $p$-$h$ symmetric point $\tilde{x}=0$ (solid). The $o$-spectrum evolves continuously as the transition is crossed, a pole occuring at the Fermi level precisely at the transition where $\epsilon_{o}^{*}$ vanishes. ~\emph{Left inset}: $e$-orbital spectrum 
$2\pi\Gamma D_{ee}(\omega)$ \emph{vs} $\omega/\Gamma$ just on either side of the QPT, for
$\tilde{x}=12.261351440$ in the FL phase (solid line) and $\tilde{x}=12.261351420$ in the USC phase (dashed). Here the entire spectrum changes abruptly on crossing the transition.
}
\end{figure}

  As a brief illustration of single-particle dynamics along the $y=x$ line, \fref{fig:fig15} 
shows the evolution of the $o$-orbital spectrum $D_{oo}(\omega) =-\tfrac{1}{\pi}\mathrm{Im}G_{oo}(\omega)$
on decreasing $\tilde{x}$ through the FL phase (main panel), across the transition and into the USC phase (right inset). In the vicinity of the QPT coming from the FL side, a strong low-frequency spectral resonance  (for $\omega >0$) is seen to develop; becoming a pole at the Fermi level precisely at the transition, and crossing  
smoothly to $\omega <0$ in the USC phase. The position of the resonance tracks the vanishing renormalized
level $\epsilon_{o}^{*}$ (\eref{eq:olevelvanish}), the $\omega =0$ pole at the transition reflecting
$\epsilon_{o}^{*}=0$ (from
\eref{eq:Goo}, using $\Sigma_{oo}(\omega=0) \equiv \Sigma_{oo}^{R}(0)$ together with \eref{eq:renormeo},
the Fermi level spectrum is given generally by $D_{oo}(\omega =0) =\delta(\epsilon_{o}^{*})$). In the vicinity
of the transition, the renormalized level $\epsilon_{o}^{*}$ is the counterpart 
of the low-energy scale $T_{*}$ introduced in \sref{sec:firstorderthermo} in respect of thermodynamics
(see \eg \fref{fig:fig6}); $T_{*}$ and $\epsilon_{o}^{*}$ both vanishing linearly as the transition is approached, and controlling the low-energy behavior of appropriate thermodynamics and single-particle dynamics respectively.

 We also add that, as expected on physical grounds, the vanishing $o$-orbital renormalized level does not show up in the corresponding $e$-channel spectrum $D_{ee}(\omega)$, which as illustrated in \fref{fig:fig15} (left inset)
changes in a wholly discontinuous fashion on crossing the transition; commensurate with the inherently first-order nature of the transition along the $y=x$ line.


\section{Experiment}
\label{sec:exp}

  We now consider the experiments of Kogan \emph{et al}~\cite{Kogan} on a GaAs-based single-electron 
transistor at low temperature ($T$), embodied in the differential conductance as a function of gate 
voltage $\dvg$ (measured relative to a reference voltage), and also the bias (or source-drain) voltage $\vsd$.
On varying the gate voltage, the resultant conductance maps shown \eg in fig. 1 of [\onlinecite{Kogan}] (see also the theoretical \fref{fig:fig16} below) show clear zero-bias Kondo peaks arising in the centers of \emph{adjacent} Coulomb blockade valleys; one valley thus being associated with an odd number of dot electrons and the other with an even number. The former valley, which extends over a relatively wide $\dvg$ range, is 
naturally interpreted~\cite{Kogan} as the normal FL, or `singlet phase'; while the latter,
extending over a narrower $\dvg$ range, is interpreted~\cite{Kogan} as the `triplet phase' (\ie the USC phase).

  In considering theoretically the conductance, 
\begin{equation}
\label{eq:zbcs6}
\frac{G_{c}(T,\vsd =0)}{(2e^{2}/h)G_{0}}~=~\int^{\infty}_{-\infty} d\omega ~ \frac{-\partial f(\omega)}{\partial\omega}~2\pi\Gamma D_{ee}(\omega)
\end{equation}
is exact~\cite{meirwingreen} at zero-bias (as before we consider explicitly $V_{2}=V_{1}$), with $D_{ee}(\omega)$ the spectrum at the temperature of interest. At finite bias by contrast, nothing exact can be said with the methods at hand. To treat approximately $\vsd \neq 0$ we neglect explicit dependence of the self-energies on $\vsd$. With this standard approximation $G_{c} \equiv G_{c}(T,\vsd)$ is readily shown to be 
\begin{equation}
\label{eq:gcs6}
\frac{G_{c}(T,\vsd)}{(2e^{2}/h)G_{0}}~\simeq~\tfrac{1}{2}\int^{\infty}_{-\infty} d\omega ~ \left[\frac{-\partial f_{L}(\omega)}{\partial\omega}+\frac{-\partial f_{R}(\omega)}{\partial\omega}\right]~2\pi\Gamma D_{ee}(\omega)
\end{equation}
where $f_{\nu}(\omega) = f(\omega \pm \tfrac{1}{2}e\vsd )$ for lead $\nu =R/L$ respectively
($f(\omega) = [e^{\omega/T}+1]^{-1}$). For $\vsd =0$, \eref{eq:gcs6} reduces correctly
to \eref{eq:zbcs6}; while for $T=0$ it yields
\begin{equation}
\label{eq:gcs6t=0}
\frac{G_{c}(T=0,\vsd)}{(2e^{2}/h)G_{0}}~\simeq~\pi\Gamma \left[D_{ee}(\omega =\tfrac{1}{2}e\vsd)
+D_{ee}(\omega =-\tfrac{1}{2}e\vsd)\right]
\end{equation}
in terms of the $T=0$ spectra. In the above we have taken a symmetric voltage split between the
$R/L$ leads. From \eref{eq:gcs6} this gives $G_{c}(T,\vsd) =G_{c}(T,-\vsd)$, which symmetry is rather
well satisfied in experiment (fig. 1, [\onlinecite{Kogan}]).

\begin{figure}
\includegraphics{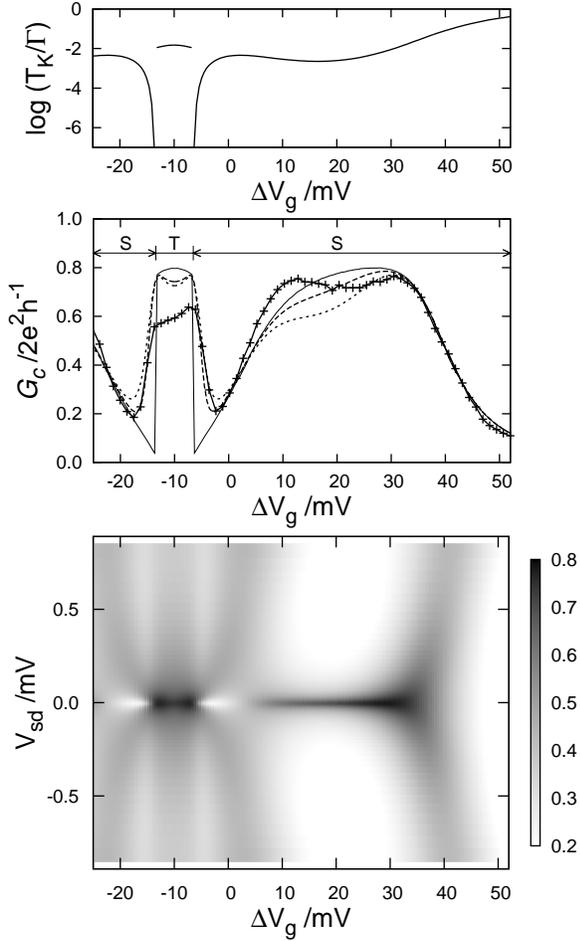}
\caption{\label{fig:fig16} 
\emph{Middle panel}: Experimental zero-bias conductance~\cite{Kogan,footnotefour} (crosses) \emph{vs} $\dvg$ (in
$\mathrm{mV}$). Theoretical results obtained as described in text are also shown, for $T/\Gamma =0$ (solid line),
$5\times 10^{-3}$ (long dash) and $10^{-2}$ (short dash); for $\Gamma = 0.5 \mathrm{meV}$ these correspond respectively to $T=0$, $30\mathrm{mK}$ and $60\mathrm{mK}$. The positions of the theoretical phase boundaries ($T=0$) between the USC triplet (T) and Fermi liquid singlet (S) phases are as indicated. \emph{Upper panel}:
Kondo scales determined as in \sref{sec:thermo}, and shown as $\log (T_{K}/\Gamma)$ \emph{vs} $\dvg$ (for the T phase, $T_{K} \equiv T_{K}^{S=1}$).
\emph{Bottom panel}: Theoretical differential conductance map in the ($\vsd, \dvg$)-plane, with the grey-scale code indicated (in units of $2e^{2}/h$). Shown for $T=30\mathrm{mK}$, choosing $\Gamma = 0.5 \mathrm{meV}$ (see text). The experimental counterpart is shown in fig.1 of [\onlinecite{Kogan}].
}
\end{figure}

  Under application of a gate voltage, the level energy $\epsilon_{1}\propto \dvg$, and one expects
the level spacing $\Delta\epsilon =\epsilon_{2}-\epsilon_{1}$ to be essentially fixed~\cite{footnotethree}. 
The experimental `trajectory' in the ($\epsilon_{1},\epsilon_{2}$) (or ($x,y$)) plane upon varying $\dvg$ is then as indicated schematically in \fref{fig:fig2}(a), \viz $y=x+\Delta\epsilon$ (\ie $\epsilon_{2}=\epsilon_{1}+\Delta\epsilon$). In this regard an interesting symmetry arises. Indicating explicitly the $x,y$ dependence of $D_{ee}(\omega) \equiv D_{ee}(\omega; x,y)$, it is readily shown that under  the $p$-$h$ and 1-2 transformations of \sref{sec:symmetries}, $D_{ee}(\omega; x,y)=D_{ee}(-\omega; -y,-x)$. Employing this in
\eref{eq:gcs6}, noting that $[\partial f_{L}(\omega)/\partial\omega +\partial f_{R}(\omega)/\partial\omega ]$
is even in $\omega$, gives
\begin{equation}
\label{eq:gcsymm}
G_{c}(T,\vsd; x,y)~=~G_{c}(T,\vsd; -y,-x)
\end{equation}
for the conductance $G_{c}(T,\vsd) \equiv G_{c}(T,\vsd; x,y)$. That is, the conductance is symmetric under reflection about the line $y=-x$. Now the `trajectory' $y=x+\Delta\epsilon$ is perpendicular to the line $y=-x$, and intersects it for $x=-\tfrac{1}{2}\Delta\epsilon$, \ie (since $x=\epsilon_{1}+\tfrac{1}{2}U+\up$)
for $\epsilon_{1}=-\tfrac{1}{2}\Delta\epsilon -\tfrac{1}{2}U-\up \equiv \epsilon_{1,\mathrm{m}}$. Since the phase boundaries are also symmetric under reflection about the line $y=-x$ (\frefss{fig:fig2}{fig:fig5}{fig:fig7}), this value $\epsilon_{1}=\epsilon_{1,\mathrm{m}}$ -- and hence the corresponding $\Delta V_{\mathrm{g,m}}$ ($\epsilon_{1} \propto \dvg$) -- is thus the midpoint of the triplet phase; such that the conductance should be an even function of $\epsilon_{1}-\epsilon_{1,\mathrm{m}}$, or equivalently of $\dvg -\Delta V_{\mathrm{g,m}}$. This symmetry is quite well satisfied in experiment (and is of course obeyed precisely in the theoretical results). 
\Fref{fig:fig16} (middle panel) shows the experimental zero-bias conductance~\cite{Kogan,footnotefour} (crosses) as a function of $\dvg$ (in $\mathrm{mV}$), together with corresponding theoretical results (as detailed below). The midpoint of the triplet (T) phase is readily identified as $\Delta V_{\mathrm{g,m}}=-10~\mathrm{mV}$, about which 
the experimental conductance is indeed seen to be quite symmetric. And the experimental conductance map shown in fig. 1 of [\onlinecite{Kogan}] (\emph{cf} \fref{fig:fig16} bottom panel) is also clearly rather symmetric about $\Delta V_{\mathrm{g,m}}$.

To compare directly to experiment we must specify the dimensionless interactions $\util, \uptil,\Jtil$ 
(\eref{eq:tildes}), $\Delta\tilde{\epsilon} =\Delta\epsilon/\Gamma$, the relation between $\tilde{\epsilon}_{1}=\epsilon_{1}/\Gamma$ and $\dvg$, and finally the hybridization strength $\Gamma$.
This is obviously a large parameter space, and our intent here is simply to employ 
what we regard as a reasonable set of `bare' parameters. 
For a typical dot the relative hierarchy of energies satisfies~\cite{PustGlazRev}
$|\Jtil| \ll \Delta\tilde{\epsilon} \ll \util$; with which the specific parameters we use here concur,
$|\Jtil| =0.5, \Delta\tilde{\epsilon}=4.5$ and $\util =12$ (and with $\util$ and $\Delta\tilde{\epsilon}$ in 
excess of unity, consistent with the occurrence of charge quantization towards the centers of the Coulomb blockade valleys). No attempt to explain experiment on the assumption $\uptil =\util$  was found to be successful, even qualitatively, on varying the bare parameters. The main reason  (as evident from inspection \eg of 
\fref{fig:fig2}(c) or \fref{fig:fig5} (top, (c)) is that the resultant width of the T phase (in 
$\tilde{\epsilon}_{1}$ or $\dvg$) is much too large compared to that of the singlet (S) phase, and as such not qualitatively consistent with experiment~\cite{Kogan} (\fref{fig:fig16}). For the results shown here we have used
$\uptil =6 =\util/2$ (although tolerable variations from this value give comparable agreement with experiment).

  From the discussion above, the relation between $\dvg$ and $\tilde{\epsilon}_{1}$ is of form
$\dvg =c[\tilde{\epsilon}_{1}-\tilde{\epsilon}_{1,\mathrm{m}}] +\Delta V_{\mathrm{g,m}}$ where the proportionality
constant $c$ is to be determined (as above, $\Delta V_{\mathrm{g,m}}=-10 \mathrm{mV}$ and
$\tilde{\epsilon}_{1,\mathrm{m}} =-\tfrac{1}{2}\Delta\tilde{\epsilon}-\tfrac{1}{2}\util -\uptil$).
For a chosen set of $\util, \uptil, \Jtil, \Delta\tilde{\epsilon}$, the theoretical zero-bias conductance
at $T=0$ is calculated from \eref{eq:zbcs6} as a function of $\tilde{\epsilon}_{1}$. It is then scaled onto the experimental results shown in \fref{fig:fig16} (middle), over the $\dvg$ range above $\sim  35-40 \mathrm{mV}$.
 We choose this range because here the system is beginning the approach to the `empty orbital' regime of 
$\nimp \ll 1$, where one does not expect any appreciable $T$-dependence to the conductance [the experimental $T$ is not known with certainty~\cite{footnotefour}, for although the experiments were performed at the refrigerator base temperature of $\sim 12 \mathrm{mK}$, the electron temperature, $T$, was not determined; although it is believed
to be $\lesssim 40 \mathrm{mK}$~\cite{footnotefour}]. With this procedure we determine the constant $c$, which
is then fixed and used for all $\dvg$ (and $T$); as well as the dimensionless constant $G_{0}$ reflecting
(\sref{sec:model}) the relative asymmetry in tunnel coupling to the leads (from scaling the vertical axis in \fref{fig:fig16}, and leading to $G_{0} \simeq 0.8$ -- as is obviously reasonable even from cursory inspection of the experimental data). 

  In comparing to experiment, an obvious key element is the relative widths (in $\tilde{\epsilon}_{1}$ or $\dvg$) of
the S and T phases, the former being considerably wider than the latter in experiment. This we naturally find to be influenced significantly by the exchange $\Jtil$ (and to a lesser extent by $\Delta\tilde{\epsilon}$), which
is optimised accordingly. For the results shown here, we find $\Jtil = -0.5$ to be optimal. Its
magnitude is small, as  expected, although its sign is antiferromagnetic. This is not however unreasonable,
for on coupling to the leads as mentioned in \sref{sec:thermo}, an AF \emph{bare} $\Jtil$ still
generates an effective ferromagnetic spin-coupling via an RKKY interaction (as evident in the very existence of
the USC triplet phase for weakly AF bare $\Jtil$, and which effect is in fact largest for the case $V_{2}=V_{1}$ we 
consider explicitly).

  With the above we calculate the $T=0$ zero-bias conductance, shown in the middle panel (solid line) of
\fref{fig:fig16} for $\util =12, \uptil =6, \Delta\tilde{\epsilon}=4.5, \Jtil =-0.5$, with the two T/S phase boundaries indicated in the figure. The T phase, symmetrically disposed about the midpoint $\Delta V_{\mathrm{g,m}}=-10\mathrm{mV}$, occurs in the interval $-13.45\mathrm{mV} \leq \dvg \leq -6.55\mathrm{mV}$
(corresponding to $-15.2 \leq \tilde{\epsilon}_{1} \leq -13.3$); with $\nimp$ at the phase boundaries of 
$\nimp =1.87$ (upper boundary at $\dvg = -6.55\mathrm{mV}$) and $\nimp =4-1.87=2.13$ (lower boundary), such that in accordance with \eref{eq:zbctwophases} of \sref{sec:zbcfsr} the zero-bias conductance increases on crossing 
from the S (FL) to the T (USC) phase. 

The resultant Kondo scales as a function of $\dvg$ are shown in the
top panel of \fref{fig:fig16} (obtained as specified in \sref{sec:thermo}, and with $T_{K} \equiv T_{K}^{S=1}$
for the T phase).  Since $T_{K}$ vanishes as the QPT is approached from the S (FL) side, finite-temperature
effects will obviously be most significant in the vicinity of the transition. \Fref{fig:fig16} thus shows
the zero-bias conductance at two non-zero temperatures, $T/\Gamma = 0.005$ and $0.01$.  While there is not much net difference between the two, each has the effect of significantly increasing the conductance in the vicinity of the transition, and leads to what we regard as rather good overall agreement with experiment. Over the $T$-range shown the conductance `inside' the T phase does not erode as rapidly as one might like, reflecting the fact that $T_{K} \equiv T_{K}^{S=1}$ therein is in excess of the $T$'s shown; although one could likely improve on this with a bare parameter set for which $T_{K}^{S=1}$ inside the T phase is somewhat smaller. The temperature range considered here is also entirely reasonable in relation to the experimental $T$  discussed above~\cite{Kogan,footnotefour}; 
with $\Gamma = 0.5\mathrm{meV}$ (as employed below) the temperatures shown correspond to 
$T=30\mathrm{mK}$ and $60\mathrm{mK}$ respectively.

 
\subsection{Conductance maps}
\label{sec:condmaps}  

The bare parameters specified above are fixed. Using \eref{eq:gcs6} the differential conductance, as a 
function of gate  \emph{and} bias voltages, may now be calculated and compared to experiment. 
For this we must finally specify the hybridization strength $\Gamma$; in the following we take $\Gamma = 0.5\mathrm{meV}$ (noting that comparison to experiment is not critically dependent on this choice, 
with values in the range $\sim 0.3-0.6\mathrm{meV}$ being found quite acceptable).
The resultant differential conductance map for $T=30\mathrm{mK}$ is shown in \fref{fig:fig16} (bottom
panel), and is in rather good agreement with the experimental results reported in fig. 1 of
[\onlinecite{Kogan}].

  In addition to the clear zero-bias Kondo ridges associated with both the T and S phases, the conductance
map shows other features noted in experiment~\cite{Kogan}. In particular, looking at
the far left side of the conductance plot, one sees two dark `ridges' positioned symmetrically around $\vsd =0$. As 
$\dvg$ is increased the two ridges move together, until they merge to form the zero-bias Kondo ridge associated with
the T phase. The latter persists for a range of $\dvg$, and then the two ridges separate again (the pattern 
being in other words symmetrical about $\Delta V_{\mathrm{g,m}}=-10\mathrm{mV}$, for the reasons explained 
following \eref{eq:gcsymm}). The obvious question arises as to the origin of these `ridges', which we now
consider.

  As seen in \fref{fig:fig16}, the $T=0$ zero-bias conductance increases on passing from the S (FL) to 
the T (USC) phase; and in \sref{sec:antires} we showed that such behavior was indicative of a vanishing
Kondo antiresonance in the single-particle spectrum, as the transition is approached from the S side.
\begin{figure}
\includegraphics{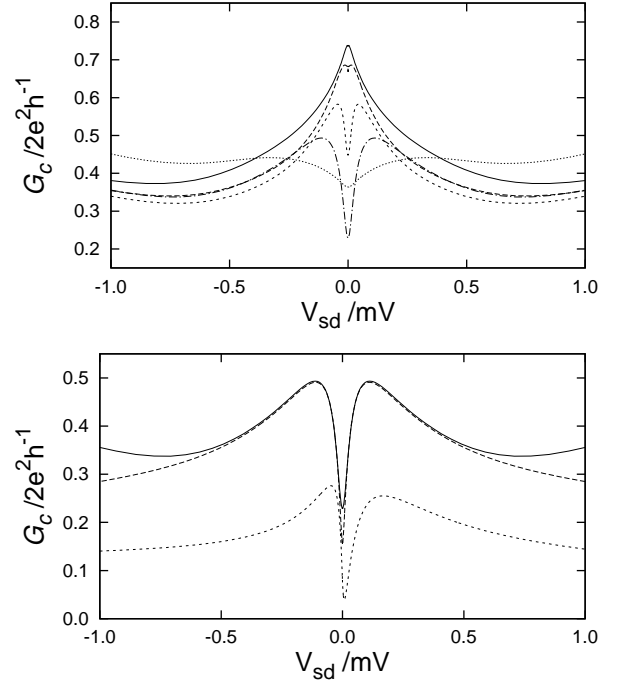}
\caption{\label{fig:fig17} 
\emph{Top panel}: 
$\vsd$-dependence of conductance, taking cuts through the conductance map of \fref{fig:fig16} (bottom) for a sequence of different fixed gate voltages: $\dvg = -10 \mathrm{mV}$ (midpoint of the T phase, sold line),
and $\dvg/\mathrm{mV} = -6$ (long dash), $-5$ (short dash), $-3$ (point dash) and $+2$ (dotted), on moving 
into and through the S phase. A clear antiresonance develops in the S phase, as discussed in text. The peaks 
in the conductance, symmetrical about $\vsd =0$, lie at the center of the ridges seen in the conductance map. \\ \emph{Bottom panel}: Focus on the $\dvg = -3\mathrm{mV}$ case. Solid line: conductance as in top panel. 
Long dash: corresponding $T=0$ conductance, proportional to $[D_{ee}(\omega =+\tfrac{1}{2}e\vsd)+D_{ee}(\omega =-\tfrac{1}{2}e\vsd)]$. Short dash: contribution from $D_{ee}(\omega = +\tfrac{1}{2}e\vsd)$ alone, showing
a clear Kondo antiresonance in the single-particle spectrum itself.
}
\end{figure}
This is the origin of the ridges seen in the conductance maps, as now shown by considering cuts through
the conductance map of \fref{fig:fig16}, for a sequence of different fixed gate voltages $\dvg$.
The top panel in \fref{fig:fig17} accordingly shows the conductance \emph{vs} bias voltage $\vsd$,
for five different values of $\dvg$: $-10\mathrm{mV}$ (at the midpoint of the T phase), and
$\dvg = -6, -5, -3$ and $+2$ $\mathrm{mV}$ as one moves into the S phase (which at $T=0$ occurs for
$\dvg \geq -6.5\mathrm{mV}$). For $\dvg =-10\mathrm{mV}$ the conductance naturally peaks at $\vsd =0$.
But on entering the S phase a clear antiresonance in the conductance is seen to develop, just setting
in by $\dvg = -6\mathrm{mV}$ and deepening progressively as $\dvg$ is increased in the S phase towards $-3\mathrm{mV}$ (then naturally disappearing as one gets considerably further into the S phase, 
as illustrated by the $\dvg = +2\mathrm{mV}$ example). And the peaks in these conductance profiles, 
symmetrically disposed about $\vsd =0$, lie at the center of the ridges in the conductance map.

  To show that the above behavior indeed reflects a Kondo antiresonance in the single-particle spectrum 
itself, the bottom panel in \fref{fig:fig17} focusses on the $\dvg = -3\mathrm{mV}$ example. The solid line again
gives the conductance shown in the top panel; while the long dash line shows the corresponding $T=0$ conductance
obtained from \eref{eq:gcs6t=0}, and thus being proportional to
$[D_{ee}(\omega =+\tfrac{1}{2}e\vsd)+D_{ee}(\omega =-\tfrac{1}{2}e\vsd)]$. The latter clearly captures well
the former (the differences naturally being due to thermal smearing). Because the conductance is proportional
to the \emph{symmetrized} spectra at $\omega = \pm \tfrac{1}{2}e\vsd$, it is not \emph{a priori} clear that the
single-particle spectrum itself contains a Kondo antiresonance. That it does, however, is seen from 
short-dash line in \fref{fig:fig17}(bottom), which shows the contribution from 
$D_{ee}(\omega = +\tfrac{1}{2}e\vsd)$ itself, seen to contain a clear Kondo antiresonance centered on the Fermi level. We add too that, since the peaks/ridges in the conductance stem from the `peaks' inherent to the Kondo antiresonance in $D_{ee}$, they are obviously not interpretable in terms of isolated dot states.

  \begin{figure}
\includegraphics{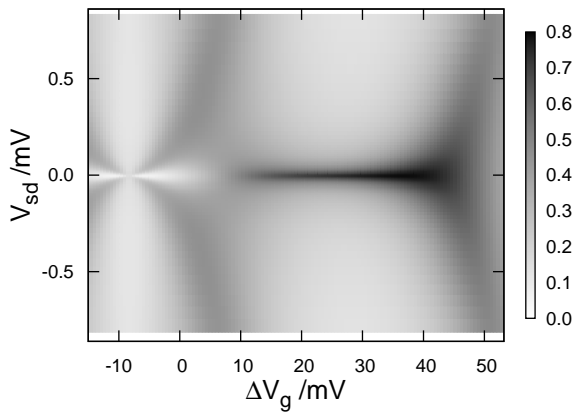}
\caption{\label{fig:fig18} As discussed in text, differential conductance map in the ($\vsd,\dvg$)-plane; shown for $T=30\mathrm{mK}$ with the same bare parameters as \frefs{fig:fig16}{fig:fig17}, except for a slight change in $\Jtil$. In this case no S/T transition occurs, the system remaining throughout in the S phase. The zero-bias Kondo ridge associated with the T phase is thus absent, and instead a zero-bias conductance antiresonance persists. 
}
\end{figure}

Finally, the zero-bias Kondo ridge in the conductance map -- formed as described above on 
merging of the $\vsd \neq 0$ conductance ridges, and concomitant vanishing of a conductance antiresonance (as in \fref{fig:fig17} top) -- reflects of course the existence of the T (USC) phase, and hence a transition to it from the S (FL) phase.
However one can readily envisage a situation where the underlying bare parameters of the system/device are
slightly different, such that on ramping down the gate voltage the resultant trajectory $y=x+\Delta\epsilon$ comes close to but `misses' the S/T transition; the system as such always remaining in a S phase (see \eg \fref{fig:fig2}(a)). In this case no zero-bias Kondo ridge associated with the T phase can arise. 
Instead,  from the discussion above, one might intuitively expect continued persistence of the conductance 
antiresonance on decreasing $\dvg$, with attendant finite-bias conductance ridges which never quite merge together.
That this indeed occurs is illustrated in \fref{fig:fig18}, where the conductance map (here for $T=30\mathrm{mK}$) is shown for the same bare parameters as \frefs{fig:fig16}{fig:fig17}, except for a slight change in $\Jtil$ to $-0.6$ (the same behavior arising also on changing \eg $\Delta\tilde{\epsilon}$ rather than $\Jtil$).
And the qualitative behavior seen here is indeed similar to that observed in a second device, 
shown in fig. 3 of [\onlinecite{Kogan}] (although in this case we have not made a quantitative comparison).


\section{Concluding remarks}

  As exemplars of multilevel quantum dot systems, we have considered in this paper correlated two-level quantum dots, coupled in a 1-channel fashion to metallic leads.  Thermodynamics, single-particle dynamics and electronic transport properties show the physical behavior of the system to be rich and varied; and our aim
has been to obtain a unified understanding of the problem for essentially arbitrary dot charge/occupancy.
Excepting points of high symmetry where first order level-crossing transitions arise, 
associated quantum phase transitions are 
of Kosterlitz-Thouless type, evident in a vanishing Kondo scale as the transition to the underscreened spin-1 phase is approached from the Fermi liquid side; and manifest in particular by a discontinuous jump in the zero-bias conductance as the transition is crossed, which we have shown can be understood here from an underlying Friedel-Luttinger sum rule.
We add in fact that an abrupt conductance change appears to be a general signature of a KT transition, such behavior arising generally not only in the present model, but also 
in capacitively coupled
2-channel quantum dots which exhibit a KT transition from a charge-Kondo Fermi liquid state (with a quenched charge pseudospin) to a non-Fermi liquid, doubly degenerate
`charge-ordered' phase~\cite{ddprl};  and in the problem of spinless, capacitively coupled metallic islands/large dots close to a degeneracy point between $N$ and $N+1$ electron states, described by two Ising-coupled Kondo impurities~\cite{garst}.



  Several 
issues 
naturally remain to be addressed. We believe for example that the generalization of Luttinger's theorem to the singular Fermi liquid USC phase (\sref{sec:FLsumrule}) is significant, and raises 
important basic questions (such as why, and what fundamentally does it reflect?). While we do not doubt its validity, 
we have however demonstrated it only numerically; and a proper analytical understanding of the result is obviously desirable.
In this work we have also considered the system in the absence of an applied magnetic field, $B$. 
Interesting physics arises also for $B\neq 0$ (see \eg [\onlinecite{pustborda}]), where the underlying quantum phase transitions are naturally smeared into crossovers. In fact, for the USC phase the limits of zero field and
$B \rightarrow 0+$ are different for $T=0$, reflecting the total polarization of a free spin-1/2 (as for the USC fixed point) on application of even an infinitesimal field.  We will turn in subsequent work to the effects of magnetic fields upon single-particle dynamics and transport in the model.


\begin{acknowledgments}
Helpful discussions with L. Borda, F. Essler, D. Goldhaber-Gordon, A. Kogan, A. Mitchell and M. Pustilnik  are gratefully acknowledged. Particular thanks are due to A. Kogan and D. Goldhaber-Gordon for kindly providing us with their experimental data from~[\onlinecite{Kogan}].
We thank the EPSRC for financial support, under grant EP/D050952/1.
\end{acknowledgments}

\appendix
\section{Effective low-energy models}
\label{sec:appendixA}
We first sketch the derivation of the effective low-energy model considered in \sref{sec:AFJ}, spanned
by the ($1,1$) triplet \emph{and} ($1,1$) singlet states of the isolated dot; with energies
under $\Hdot$ of $E_{T}=\epsilon_{1}+\epsilon_{2}+\up -\tfrac{1}{4}\J$ and
$E_{S}=\epsilon_{1}+\epsilon_{2}+\up +\tfrac{3}{4}\J$ respectively. The local unity operator
for the dot is 
\begin{equation}
\label{eq:unity}
\hat{1}~=~\hat{1}_{\mathrm{T}}~+~\hat{1}_{\mathrm{S}}
\end{equation}
with $\hat{1}_{\mathrm{T}}=\sum_{S^{z}} ~|S=1, S^{z}\rangle\langle S=1, S^{z}|$ in an obvious notation, and
likewise $\hat{1}_{\mathrm{S}}=|0,0\rangle\langle 0,0|$. These satisfy
the following identities in the local Hilbert space,
\begin{equation}
\label{eq:identities}
\hat{1}_{\mathrm{T}}~=~~\hat{\mathbf{s}}_{1}\cdot\hat{\mathbf{s}}_{2}~+~\tfrac{3}{4}\hat{1} ~~~~~~~~~~
\hat{1}_{\mathrm{S}}~=~-\hat{\mathbf{s}}_{1}\cdot\hat{\mathbf{s}}_{2}~+~\tfrac{1}{4}\hat{1},
\end{equation}
as follows using $\hat{\mathbf{s}}_{1}\cdot\hat{\mathbf{s}}_{2} \equiv \half\left(\hat{S}^{2} -\tfrac{3}{2}\right)$
with $\hat{\mathbf{S}}=\hat{\mathbf{s}}_{1}+\hat{\mathbf{s}}_{2}$.

Omitting for brevity the lead contribution $\Hl$ (\eref{eq:hl}), the low-energy model is given  by
$\hat{H}_{\mathrm{eff}} = E_{S}\hat{1}_{\mathrm{S}}+E_{T}\hat{1}_{\mathrm{T}} +\hat{H}^{(2)}_{\mathrm{eff}}$.
The first two terms are simply the bare energies of the dot states; using 
\erefs{eq:unity}{eq:identities} they may be written as
$E_{S}\hat{1}_{\mathrm{S}}+E_{T}\hat{1}_{\mathrm{T}} =\tfrac{1}{4}(E_{S}+3E_{T})\hat{1} +(E_{T}-E_{S})\hat{\mathbf{s}}_{1}\cdot\hat{\mathbf{s}}_{2}$, or equivalently as
$-\J\hat{\mathbf{s}}_{1}\cdot\hat{\mathbf{s}}_{2}$ on omitting the first (constant/common) term:
\begin{equation}
\hat{H}_{\mathrm{eff}}~=~-\J\hat{\mathbf{s}}_{1}\cdot\hat{\mathbf{s}}_{2}~+~\hat{H}^{(2)}_{\mathrm{eff}}
\end{equation}
Here $\hat{H}^{(2)}_{\mathrm{eff}}$ is the leading (${\cal{O}}(V^{2})$) contribution arising from tunnel coupling to
the leads (\eref{eq:ht} with $V_{2}=V_{1} \equiv V$, here denoted as $\hat{H}^{\prime}$), given from a
SW transformation~\cite{schriefferwolff} as
\begin{equation}
\label{eq:SW}
\hat{H}^{(2)}_{\mathrm{eff}}=\tfrac{1}{2} \sum_{\alpha,\beta} ~\hat{1}_{\beta} \hat{H}^{\prime} \left[ \left(E_{\alpha}-\hat{H}_{D}\right)^{-1} +\left(E_{\beta}-\hat{H}_{D}\right)^{-1}\right] \hat{H}^{\prime}\hat{1}_{\alpha}
\end{equation}
with $\alpha,\beta \in \{S,T\}$ (retardation effects are as usual neglected).

In analyzing \eref{eq:SW} one encounters the following `natural' exchange couplings $J_{i}^{\alpha} >0$,
\begin{equation}
\label{eq:Jialpha}
J_{i}^{\alpha}~=~NV^{2} \left[ \frac{1}{\Delta E_{i}^{\alpha}} +
\frac{1}{\Delta \tilde{E}_{i}^{\alpha}} \right]
\end{equation}
(with $N$ the number of $\K$-states in the lead, such that $NV^{2} \sim {\cal{O}}(1)$).
Here the $\Delta E_{i}^{\alpha}>0$ denote electron \emph{removal} excitation energies from level $i =1$ or $2$,
relative to the $\alpha =T$ or $S$ dot ground state, and the
$\Delta \tilde{E}_{i}^{\alpha}>0$ correspondingly denote electron \emph{addition} energies to level
$i$ relative to the $\alpha = T$ or $S$ ground state; \eg $\Delta E^{T}_{1} =E_{D}(0,1)-E_{T}$ or
$\Delta\tilde{E}^{S}_{2} =E_{D}(1,2)-E_{S}$ in the notation of \sref{sec:phaseoverview}.
Denoting
\begin{equation}
\label{eq:lambdaSandT}
\lambda_{T} =\half U +\tfrac{1}{4}\J ~~~~~~~~  \lambda_{S} =\half U -\tfrac{3}{4}\J
\end{equation}
these excitation energies are easily shown to be given by
\begin{subequations}
\begin{equation}
\Delta E_{1}^{\alpha} = \lambda_{\alpha}- x ~~~~~~~~~ \Delta \tilde{E}_{1}^{\alpha} = \lambda_{\alpha}+ x
\end{equation}
\begin{equation}
\Delta E_{2}^{\alpha} = \lambda_{\alpha}- y ~~~~~~~~~ \Delta \tilde{E}_{2}^{\alpha} = \lambda_{\alpha}+ y
\end{equation}
\end{subequations} 
where (\eref{eq:xandy}) $x=\epsilon_{1}+\half U +\up$, $y=\epsilon_{2}+\half U +\up$. Hence from
\eref{eq:Jialpha},
\begin{equation}
J_{1}^{\alpha} \equiv J^{\alpha}(x) ~~~~~~~~ 
J_{2}^{\alpha} \equiv J^{\alpha}(y)
\end{equation}
with $J^{\alpha}(x)$ defined by:
\begin{equation}
\label{eq:Jalpha}
J^{\alpha}(x)~=~ NV^{2} \left[ \frac{1}{\lambda_{\alpha}-x} ~+~\frac{1}{\lambda_{\alpha}+x}\right] ~~~~ = ~J^{\alpha}(-x)
\end{equation}

 Direct analysis of \eref{eq:SW} yields, after a standard if laborious calculation, the effective low-energy
model \eref{eq:heff},
\begin{equation}
\label{eq:heffapp}
\hat{H}_{\mathrm{eff}}~=~ J_{1}(x,y)~\hat{\mathbf{s}}_{1}\cdot\hat{\mathbf{s}}_{0}~+~
J_{2}(x,y)~\hat{\mathbf{s}}_{2}\cdot\hat{\mathbf{s}}_{0}~-~I(x,y)~\hat{\mathbf{s}}_{1}\cdot\hat{\mathbf{s}}_{2}
\end{equation}
where potential scattering contributions are omitted for clarity, and
$\hat{\mathbf{s}}_{0}$ denotes the spin density of the conduction channel at the dot.
The direct exchange coupling between spins $1$ and $2$ is found to be
\begin{equation}
\label{eq:Idirect}
I(x,y)~=~ \J ~+~ \tfrac{1}{2} \left[ J^{T}(x)+J^{T}(y)-J^{S}(x) -J^{S}(y)\right]
\end{equation}
while the ${\cal{O}}(V^{2})$ antiferromagnetic exchange couplings between spins $1$ or $2$ and the
lead are given by
\begin{subequations}
\label{eq:JoneJtwo}
\begin{equation}
\label{eq:Jone}
J_{1}(x,y)~=~\tfrac{1}{2} \left[ 3J^{T}(x)+J^{T}(y) +\left(J^{S}(x)-J^{S}(y)\right)\right]
\end{equation}
\begin{equation}
\label{eq:swap}
J_{2}(x,y)~=~J_{1}(y,x).
\end{equation}
\end{subequations}
Using \eref{eq:Jalpha}, these exchange couplings satisfy 
\begin{subequations}
\begin{equation}
\label{eq:Jisymm}
J_{i}(x,y)~=~J_{i}(-x,-y)~~~~~~~~: i=1,2 
\end{equation}
\begin{equation}
I(x,y)~~=~~I(-x,-y) ~~=~~ I(y,x)~~~~
\end{equation}
\end{subequations}
by virtue of which $\hat{H}_{\mathrm{eff}}\equiv \hat{H}_{\mathrm{eff}}(x,y)$ in \eref{eq:heffapp}
satisfies $\hat{H}_{\mathrm{eff}}(x,y) \equiv \hat{H}_{\mathrm{eff}}(-x,-y)$ (reflecting its symmetry 
under a $p$-$h$ transformation); while from \eref{eq:swap}, 
$\hat{H}_{\mathrm{eff}}(x,y) \rightarrow \hat{H}_{\mathrm{eff}}(y,x)$ 
under the 1-2 transformation (\eref{eq:12t}), as expected on general grounds from \sref{sec:symmetries}.
These symmetries, and the consequent invariance of the phase boundaries in the ($x,y$)-plane to both inversion and reflection about the line $y=x$, are also naturally satisfied when potential scattering terms, omitted explicitly from \eref{eq:heffapp}, are included. By contrast, the apparent reflection symmetries 
$\hat{H}_{\mathrm{eff}}(x,y) = \hat{H}_{\mathrm{eff}}(-x,y) = \hat{H}_{\mathrm{eff}}(x,-y)$ 
which hold for \eref{eq:heffapp} itself (via \eref{eq:Jalpha}), are not preserved when
potential scattering is included; which is why the phase boundaries in \fref{fig:fig7} are not
invariant to reflection about the lines $x=0$ and $y=0$.

  Along the lines $y=\pm x$ in the ($x,y$)-plane, it follows directly from \erefs{eq:swap}{eq:Jisymm}
that $J_{2}(x,\pm x)=J_{1}(x,\pm x)$, and hence from \erefs{eq:Jone}{eq:Jalpha} 
\begin{equation}
\label{eq:Jtwo=Jone}
J_{2}(x,\pm x)~=~J_{1}(x,\pm x)~=~2J^{T}(x),
\end{equation}
with $I(x,\pm x)=\J +[J^{T}(x)-J^{S}(x)]$ following similarly from \eref{eq:Idirect}. In consequence,
as discussed in \sref{sec:AFJ}, $\hat{H}_{\mathrm{eff}}$ is separable along the lines $y=\pm x$, and
first order level-crossing transitions thus arise. Note incidentally that for given $\J <0$, both $J^{T}(x)$ 
and $I(x,\pm x)$ increase on increasing  $|x|$ from $0$, both of which act 
(see \eref{eq:heffseparable}) to favor the triplet phase; consistent with the USC phase surviving
for $|x|>0$ after it has been destroyed for $x=0$,
as indeed found in \fref{fig:fig7}.

  Finally, a byproduct of the above gives directly the pure spin-1 Kondo model 
appropriate for ferromagnetic $\J >0$ deep in the USC spin-1 phase~\cite{PustGlaz2001,Posazh2006}, where the ($1,1$) singlet dot states
are energetically irrelevant and only the $(1,1)$ triplet states need be retained. To this end simply
project into the triplet sector, $\hat{H}_{K} =\hat{1}_{\mathrm{T}}\hat{H}_{\mathrm{eff}}\hat{1}_{\mathrm{T}}$.
Writing $J_{1}\hat{\mathbf{s}}_{1}\cdot\hat{\mathbf{s}}_{0}+
J_{2}\hat{\mathbf{s}}_{2}\cdot\hat{\mathbf{s}}_{0}=$
$\half (J_{1}+J_{2})(\hat{\mathbf{s}}_{1}+\hat{\mathbf{s}}_{2})\cdot\hat{\mathbf{s}}_{0}$
$+\half(J_{1}-J_{2})(\hat{\mathbf{s}}_{1}-\hat{\mathbf{s}}_{2})\cdot\hat{\mathbf{s}}_{0}$,
 recognising that $\hat{1}_{\mathrm{T}}(\hat{\mathbf{s}}_{1}-\hat{\mathbf{s}}_{2})\hat{1}_{\mathrm{T}}=0$
and neglecting constants, \eref{eq:heffapp} gives a spin-1 Kondo model
\begin{equation}
\label{eq:spin1kondo}
\hat{H}_{K}~=~\hat{1}_{\mathrm{T}}\hat{H}_{\mathrm{eff}}\hat{1}_{\mathrm{T}}~=~
J_{K}~\hat{\mathbf{S}}\cdot\hat{\mathbf{s}}_{0}
\end{equation}
where $\hat{\mathbf{S}} \equiv \hat{1}_{\mathrm{T}}(\hat{\mathbf{s}}_{1}+\hat{\mathbf{s}}_{2})\hat{1}_{\mathrm{T}}$ 
is a pure spin-1 operator, and $J_{K}=\half(J_{1}+J_{2})$ is given from \eref{eq:JoneJtwo} by
\begin{equation}
\label{eq:Jktotspin1}
J_{K}~=~ J^{T}(x)+J^{T}(y)
\end{equation}
with $J^{T}(x)$ given explicitly by \eref{eq:Jalpha}. Along the lines $y=\pm x$ in particular, \erefss{eq:Jktotspin1}{eq:Jalpha}{eq:lambdaSandT} give
$J_{K}=2J^{T}(x)$; \ie $\rho J_{K}=\tfrac{8}{\pi} \Gamma [U+\tfrac{1}{2}\J]/([U+\tfrac{1}{2}\J]^{2}-x^{2})$
with $\rho$ here the lead density of states per conduction orbital (such that $\Gamma \equiv \pi V^{2}N\rho$).
From perturbative scaling~\cite{Hewsonbook} the spin-1 Kondo scale $T_{K}^{S=1}$ follows as
$T_{K}^{S=1} \sim D \exp(-1/\rho J_{K})$ (with the exponential dependence as usual of the essence, and
the prefactor immaterial), and \eref{eq:TKspin1ph} for $T_{K}^{S=1}$ thus results.
From NRG calculations we have confirmed explicitly that the dependence of $T_{K}^{S=1}$ 
on $\util +\tfrac{1}{2}\Jtil$ is indeed as predicted by \eref{eq:TKspin1ph}.


\section{Self-energies}
\label{sec:appendixB}

The key NRG method for calculating the self-energy~\cite{bullahewprus} is readily extended to
multilevel dots/impurities. With $\hat{H}_{I}$ the interacting part of the dot Hamiltonian, 
equation of motion techniques~\cite{Zubarev} are used to obtain the following basic equation for the retarded  
propagators $\{G_{ij}(\omega)\}$:
\begin{equation}
\label{eq:B1}
\sum_{l}\left((\omega^{+} -\epsilon_{i})\delta_{il}-\Gamma_{il}(\omega)\right)G_{lj}(\omega)~=~\delta_{ij}+
\langle\langle[\dides ,\hat{H}_{I}^{\pd}];d_{j\sigma}^{\dagger}\rangle\rangle
\end{equation}
The sum is over the dot levels ($l=1,2$ here), and $[,]$ denotes a commutator. By definition, $\hat{H}_{I} \equiv 0$ in the non-interacting limit; whence \eref{eq:B1} is of form
\begin{equation}
\label{eq:B2}
\left[\bm{G}^{0}(\omega)\right]^{-1}\bm{G}(\omega)~=~\bm{1}+\bm{F}(\omega)
\end{equation}
with $\bm{G}^{0}(\omega)$ the non-interacting propagator matrix and the elements of $\bm{F}(\omega)$ given by:
\begin{equation}
\label{eq:B3}
F_{ij}(\omega)~=~\langle\langle ~[\dides ,\hat{H}_{I}^{\pd}];d^{\dagger}_{j\sigma}~\rangle\rangle
\end{equation}
Using the Dyson equation in the form
$[\bm{G}^{0}]^{-1}=[\bm{G}]^{-1}+\bm{\Sigma}$,
\eref{eq:B2} gives directly  \eref{eq:FoverG},
\begin{equation}
\label{eq:B4}
\bm{\Sigma}(\omega)~=~\bm{F}(\omega)[\bm{G}(\omega)]^{-1}
\end{equation}
from which the self-energies are calculated directly (\sref{sec:spectra}). Combining \eref{eq:B4} with
the Dyson equation also gives $\bm{G}=\bm{G}^{0}(\bm{1}+\bm{F})$, so that \eref{eq:B4} may 
be written alternatively as
\begin{equation}
\label{eq:B5}
\bm{\Sigma}(\omega)~=~\bm{F}(\omega) \left[\bm{1}+\bm{F}(\omega)\right]^{-1}\left[\bm{G}^{0}(\omega)\right]^{-1}
\end{equation}
(which we exploit to calculate the renormalized level $\epsilon_{o}^{*}$ when considering dynamics on the $y=x$ line, \eref{eq:epsilonostar} \sref{sec:y=xdynamics}).

$\hat{H}_{I}$ is given explicitly for the present problem by the separable sum (\eref{eq:hdot}) 
$\hat{H}_{I}= U\sum_{i} \hat{n}_{i\uparrow}^{\pd}\hat{n}_{i\downarrow}^{\pd} +\up\hat{n}_{1}^{\pd}\hat{n}_{2}^{\pd}
-\J~\hat{\mathbf{s}}_{1}\cdot\hat{\mathbf{s}}_{2}$. The elements \eref{eq:B3} of $\bm{F}$  are thus linearly separable 	 as $F_{ij}=F_{ij}^{U}+F_{ij}^{\up}+F_{ij}^{J}$ (in obvious notation), and are calculated individually. Since each such term is a retarded correlation function, they are Lehmann resolvable~\cite{agdbook}, and in consequence satisfy sum rules.  Writing
$F_{ij}(\omega)=F_{ij}^{R}(\omega)-\I F_{ij}^{I}(\omega)$ (with the real/imaginary parts related by Hilbert transformation), the general sum rule is
\begin{equation}
\int^{\infty}_{-\infty}\frac{d\omega}{\pi}~F_{ij}^{I}(\omega)~=~
\langle \big\{[\dides ,\hat{H}_{I}],
d_{j\sigma}^{\dagger} \big\}\rangle
\end{equation} 
where $\{,\}$ denotes an anticommutator. Specifically, for the present problem in the absence of an applied magnetic field, it is readily shown that the diagonal-element sum rule is
\begin{equation}
\int^{\infty}_{-\infty}\frac{d\omega}{\pi}~F_{ii}^{I}(\omega)~=~
\half U\langle\hat{n}_{i}^{\pd}\rangle ~+~ \up\langle \hat{n}^{\pd}_{\bar{i}}\rangle
\end{equation} 
(where $\bar{i}$ means the opposite level to $i$); while for the off-diagonal elements:  
\begin{subequations}
\begin{align}
\int^{\infty}_{-\infty}\frac{d\omega}{\pi}~F_{ij}^{I}(\omega)~&\overset{j \neq i}=~
\left(-\up + \tfrac{3}{4}\J\right) \langle d^{\dagger}_{j\sigma}\dides\rangle \\
&=~\left(-\up + \tfrac{3}{4}\J\right) \int^{0}_{-\infty}d\omega ~D_{ij}(\omega) ~~~~~
\end{align} 
\end{subequations}
These sum rules provide a check on the accuracy of the NRG calculations, and in practice are
well satisfied.


\begin{thebibliography}{49}
\bibitem[{foo({\natexlab{a}})}]{footnoteone}
\bibinfo{note}{In practice we define the low-energy scale $T_{*}$ via
  $S_{\mathrm{imp}}(T_{*}) = 0.85$, suitably between $\ln 2$ and $\ln 3$.}
\bibitem[{foo({\natexlab{b}})}]{footnotetwo}
\bibinfo{note}{The reader will notice from \fref{fig:fig7} that the first order
  transitions along $y=-x$ occur on the `inside' intersection of this line with
  the phase boundary, but not on the outer intersection. This reflects the fact
  that \eref{eq:heff} itself is not in fact applicable in the latter region,
  constructed as it is retaining solely the ($1,1$) singlet and triplet states
  of the dot. On progressively increasing $y$ along the line $y=-x$, the
  ($2,0$) (or ($0,2$)) singlet states of the dot must also be retained in the
  low-energy manifold, leading to a different effective low-energy model with a
  non-separable Hilbert space.}
\bibitem[{foo({\natexlab{d}})}]{footnotethree}
\bibinfo{note}{Any variation of $\Delta\epsilon$ with $\dvg$ is a much weaker
  effect and thus usually neglected. We can in principle include it, but in
  comparing to experiment here do not find it necessary to do so.}
\end{thebibliography}

\end{document}